\documentclass[aps,prx,twocolumn,showpacs,floatfix,superscriptaddress,tightenlines,amsmath,amssymb,longbibliography]{revtex4-2}

\usepackage[T1]{fontenc}
\usepackage{graphicx}
\usepackage{amscd}
\usepackage{amsmath,amsfonts,amssymb,mathrsfs}
\usepackage{gensymb}
\usepackage{dcolumn}
\usepackage{bm}
\usepackage{float}
\usepackage[pdftex]{color}
\usepackage{hyperref}
\hypersetup{hidelinks,breaklinks,colorlinks,citecolor=blue,linkcolor=red,urlcolor=blue}
\usepackage{textcomp}
\usepackage{array}
\usepackage{booktabs}
\usepackage{xspace}

\graphicspath{{figures/}}

\newcommand{\cns}{{CoNb$_3$S$_6$}\xspace}
\newcommand{\cts}{{CoTa$_3$S$_6$}\xspace}
\newcommand{\peakA}{{$(\tfrac{1}{2}00)$}\xspace}
\newcommand{\peakB}{{$(0\tfrac{1}{2}0)$}\xspace}

\def \angIP#1{{\mu_{#1}}}
\def \angOP#1{{\nu_{#1}}}

\let\oldcdot=\cdot
\def\cdot{\negthinspace\,\oldcdot\negthinspace\,}

\begin{document}

\title{\texorpdfstring{Double-$Q$ chiral stripe order in the anomalous Hall antiferromagnet \cns}{}}

\author{Ben Zager}
\affiliation{Department of Physics, Brown University, Providence, Rhode Island 02912, USA}

\author{Shang-Shun Zhang}
\affiliation{ Department of Physics and Astronomy, University of Tennessee, Knoxville, TN 37996, USA}

\author{Hana Schiff}
\affiliation{Department of Physics and Astronomy, University of California, Irvine, California
92697, USA}

\author{Raymond Fan}
\affiliation{Diamond Light Source Ltd., Harwell Science and Innovation Campus, Didcot OX11 0DE, United Kingdom}

\author{Paul Steadman}
\affiliation{Diamond Light Source Ltd., Harwell Science and Innovation Campus, Didcot OX11 0DE, United Kingdom}

\author{Cristian D. Batista}
\affiliation{Department of Physics and Astronomy, University of Tennessee, Knoxville, TN 37996, USA}
\affiliation{Neutron Scattering Division, Oak Ridge National Laboratory, Oak Ridge, TN 37831, USA}

\author{Kemp W. Plumb}
\email[Corresponding author: ]{kemp_plumb@brown.edu}	
\affiliation{Department of Physics, Brown University, Providence, Rhode Island 02912, USA}

\date{\today}

\begin{abstract}
We present fine momentum space resolution resonant elastic x-ray scattering measurements of the magnetic structure of the metallic antiferromagnet \cns. Using circular dichroism and full linear polarization analysis of the magnetic scattering, we reveal a non-coplanar double-$\bm{Q}$ ($2Q$) order, with a non-collinear commensurate component and a long-wavelength incommensurate helical component. This $2Q$ structure exhibits a staggered scalar spin chirality that forms a modulated stripe-like pattern with no uniform component. This novel magnetic order is naturally explained by the presence of four-spin exchange interactions and exhibits a complex domain structure that suggests a lowering of the structural symmetry. A symmetry analysis indicates that the $2Q$ order enables a finite anomalous Hall effect in \cns. In addition to identifying a novel type of magnetic ordering and its origin, our results provide insight into the mechanism of the unconventional magnetotransport phenomena in \cns and thus identifies potential routes for realizing novel electronic phenomena in metallic antiferromagnets.
\end{abstract}

\maketitle

\section{Introduction}\label{section:intro}

Competing interactions in frustrated magnets often suppress magnetic ordering, leading to highly degenerate ground states in the classical limit that evolve into quantum spin liquids when quantum fluctuations are considered. These extreme quantum states of matter, driven by frustration and that do not break symmetries, contrast with the complex multi-sublattice ordered classical magnetic structures that can also emerge from frustrated exchange interactions, exhibiting a diverse range of phenomena depending on the symmetries they break~\cite{sang2022magnetic,nakatsuji2022topological,smejkal2022beyond,smejkal2022emerging}.

For instance, the real-space Berry curvature associated with a uniform scalar spin chirality in non-coplanar magnets acts as a fictitious magnetic field in the double-exchange limit of a Kondo lattice model~\cite{Ye1999BerryPhase, taguchi2001Chirality,nagaosa2010anomalous}. Remarkably, this fictitious field, which has a uniform component even without uniform magnetization, can produce a large anomalous Hall effect (AHE) in metallic magnets~\cite{Ye1999BerryPhase,nagaosa2006anomalous}. Given that frustrated exchange and four-spin interactions often stabilize non-coplanar magnetic structures~\cite{batista2016frustration}, metallic frustrated magnets offer significant potential for fast, low-dissipation electronic and spintronic devices. This has driven substantial efforts to identify magnetic materials with uniform scalar chirality~\cite{kurumaji2019skyrmion,hirschberger2019skyrmion,takagi2023spontaneous,park2023tetrahedral}.


\begin{figure}[t!]
   \includegraphics{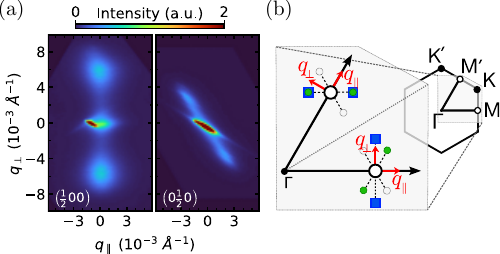}
   \caption{
   (a) Magnetic REXS intensity in S3 at 16 K. (b) Experimental geometry, incident (outgoing) polarization channels $\sigma$ ($\sigma'$) and $\pi$ ($\pi'$) correspond to $\alpha$ ($\eta$) of $0\degree$ and $90\degree$ respectively. (c) Summary of observed magnetic peaks in the triangular lattice Brillouin zone. White circles are $\bm{Q}_0\!=\! (\tfrac{1}{2}00)$ and $(0\tfrac{1}{2}0)$, green circles show the magnetic reflections observed in S1 and S2, and blue squares show those found in S3 as in (a). Dashed lines with empty markers show positions of symmetry allowed peaks that were not observed.
   }
   \label{fig:fig1}
\end{figure}


Recently, a large AHE in \cts was discovered to be associated with a uniform scalar spin chirality produced by a tetrahedral triple-$\bm{Q}$ magnetic order \cite{takagi2023spontaneous, park2023tetrahedral}. \cts belongs to a family of transition metal dichalcogenides (TMDs) intercalated with $3d$ transition metal ions. These metallic magnets exhibit diverse phenomena depending on the host compound, intercalation species, and intercalation ratio \cite{parkin19803dI,xie2022structure}. The interplay of localized spins on the $3d$ sites and itinerant electrons in the host layers, along with competing interactions gives rise to e.g., a chiral soliton lattice \cite{togawa2012chiral}, electric magnetochiral effect \cite{aoki2019nonrecip}, spin glass phenomena \cite{maniv2021exchange}, or magnetically mediated electric switching \cite{nair2020electrical}. \cns is of particular interest because it exhibits a large AHE together with a nearly vanishing uniform magnetization, implicating a non-trivial magnetic structure \cite{ghimire2018large,tenasini2020giant,parkin1983magnetic}, similar to \cts. A series of neutron diffraction measurements have found the symmetry-related magnetic propagation vectors \peakA, \peakB, and $(\tfrac{1}{2}\overline{\tfrac{1}{2}}0)$, but disagree on the orientation of the moments and the presence of single-$\bm{Q}$ ($1Q$) domains or multi-$\bm{Q}$ order \cite{parkin1983magnetic,tenasini2020giant,lu2022understanding,takagi2023spontaneous,zhang2023chiral,popcevic2023electronic}. Elucidating the precise details of the magnetic structure is an essential step towards understanding the microscopic mechanisms of symmetry breaking, the origin of the giant AHE in this material, and potentially tuning the properties to realize new functionalities.

In this work, we reexamine the magnetic structure of \cns using resonant elastic x-ray scattering (REXS) at the Co $L_3$ edge. The exceptional momentum resolution of our measurements reveals a previously undetected double-$\bm{Q}$ ($2Q$) magnetic order comprised of a non-collinear commensurate $\bm{Q}_0=(\tfrac{1}{2}00)$ component and incommensurate $\bm{Q}_0\pm\bm{q}$ helical modulation. This non-coplanar magnetic structure gives rise to a staggered ordering of scalar spin chirality with a modulated stripe or checkerboard pattern. We show that magnetic frustration generated by four-spin exchange interactions can naturally explain the observed incommensurate helical modulation \cite{heinonen2022magnetic,hayami2017effective,hayami2020multiple}. Finally, we discuss the role of the $2Q$ order in the AHE and show that the complex domain structure observed in our experiments is consistent with a lowering of the structural symmetry. Our results show that the AHE in \cns does not arise from a uniform scalar chirality as in \cts. However, in both materials four-spin interactions are crucial to stabilize the magnetic structures that break symmetries otherwise forbidding an AHE. Together, these materials demonstrate how magnetic frustration arising from four-spin interactions can in general give rise to a large electronic response in metallic antiferromagnets.

Our findings are presented as follows. In section \ref{section:methods} we describe the single crystal growth, characterization, and REXS experimental methods. The results of our REXS experiments are presented section \ref{section:results} where we first present our finding of double-$\bm{Q}$ magnetic order in section \ref{subsection:2QmagOrder}, followed by REXS circular dichroism and full linear polarization analysis to constrain the magnetic structure in sections \ref{subsection:CircDi} and \ref{subsection:FLPA}. The scalar chiral ordering is presented in section \ref{subsection:ScalarChi}. Section \ref{section:small-q} contains an explanation of the origin the observed double-$\bm{Q}$ chiral stripe order from four-spin exchange interactions. In section~\ref{section:AHE} we present symmetry arguments that connect the observed helically modulated magnetic structures with the AHE in \cns. Based on our symmetry analysis and observations of a complex magnetic domain structure, we propose that there is an, as of yet undetected, structural symmetry breaking in \cns that would provide a consistent explanation for all of our measurements. Finally, we conclude with a brief discussion in section~\ref{section:discussion}.

\section{Methods}\label{section:methods}

Single crystals were grown using chemical vapor transport \cite{ghimire2018large} with the nominal stoichiometry Co:Nb:S$=$1:3:6. Five different samples from the same growth were measured. All samples undergo magnetic transitions at 28.6~K and exhibit sharp (100) Bragg peaks, indicating a well-ordered triangular lattice of intercalated Co ions \cite{Note1,ueno2005xray}. Additional thermodynamic and transport measurements are presented in the supplementary material \cite{Note1}. REXS experiments were performed at the I10 beamline at Diamond Light Source using the RASOR endstation \cite{beale2010rasor} with the experimental geometry shown in Fig.~\ref{fig:fig1}(b). Samples were mounted in the $(HK0)$ scattering plane to access $(\tfrac{1}{2}00)$ and $(0\tfrac{1}{2}0)$ magnetic wavevectors at $2\theta\! =\! 106.2\degree$ at the Co $L_3$ edge (778.5 eV). In this geometry, the x-ray beam scatters from a natural facet on the crystal side [Fig.~\ref{fig:flpa}(a)] probing an effective area of $20\!\times\! 200\;\mu$m with a penetration depth of 0.3~$\mu$m. Thus, our measurements probe a macroscopic sample volume containing many basal plane layers, but can only access a single M and M$'$ point in reciprocal space. This experimentally accessible reciprocal space region is shown as the gray shaded area in Fig.~\ref{fig:fig1}(b). Reciprocal space maps around the $(\tfrac{1}{2}00)$ and $(0\tfrac{1}{2}0)$ magnetic Bragg peaks were collected from four different (S1-S4) samples using an area detector with $\pi$-polarized x-rays. Full linear polarization analysis (FLPA) was carried out on a fifth sample (S5) using a point detector and multilayer polarization analyzer optimized for the Co $L_3$ edge \cite{wang2012complete}. We additionally collected reciprocal space maps on S3 using circularly polarized x-rays under both zero-field cooled (ZFC) and field cooled (FC) conditions, in which the sample was cooled from far above $T_N$ with a 0.1~T magnetic field along the sample $c$-axis.

\section{Experimental Results}\label{section:results}

\subsection{\texorpdfstring{$2\bm{Q}$}{} magnetic order}\label{subsection:2QmagOrder}

Representative reciprocal space maps of the magnetic scattering observed in \cns are shown in Fig.~\ref{fig:fig1}(a). Primary magnetic reflections at $\bm{Q}_0\! =\! (\tfrac{1}{2}00)$ and $(0\tfrac{1}{2}0)$ were observed in all samples, consistent with previous reports \cite{parkin1983magnetic,tenasini2020giant,lu2022understanding,takagi2023spontaneous,zhang2023chiral}. Our fine resolution measurements also revealed new satellite magnetic reflections at $\bm{Q}_0\!\pm\!\bm{q}$ (Fig.~\ref{fig:fig1}). These satellites represent a long-wavelength incommensurate modulation of the magnetism that was not previously observed because of the relatively coarse momentum resolution of the reported neutron experiments \cite{parkin1983magnetic,tenasini2020giant,lu2022understanding, takagi2023spontaneous,zhang2023chiral}.

\begin{figure}
   \includegraphics{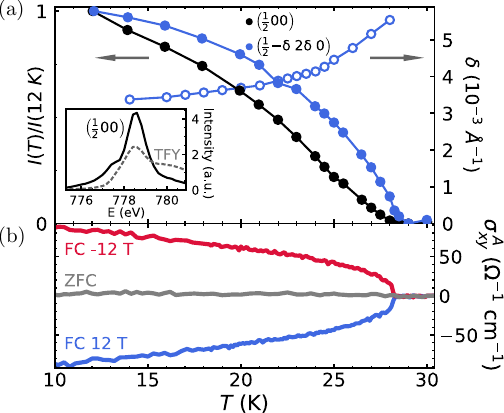}
   \caption{(a) Normalized temperature dependence of main and satellite peak, and $\delta$ in satellite wavevector $(\delta,-2\delta,0)$ for S3 measured on the point detector with $\pi$ polarized x-rays. The inset shows a fixed-$\bm{Q}$ energy scan of the main peak, with the dashed line showing total fluorescence yield (TFY). (b) Anomalous Hall conductivity $\sigma^A_{xy}$ of S3 determined from the Hall and longitudinal resistivities measured at 0 T after zero-field cooling (ZFC) and $\pm12$~T field cooling (FC).}
   \label{fig:fig2}
\end{figure}

Two different types of satellite wavevectors were observed across these samples as summarized in Fig.~\ref{fig:fig1}(b). In S1 and S2, satellites appear at $(\tfrac{1}{2},\pm\delta,0)$ and $(\pm\delta,\tfrac{1}{2},0)$ \cite{Note1}. In S3, one set of satellites appears at $(\pm\delta,\tfrac{1}{2},0)$ while the other set appears at $(\tfrac{1}{2}\mp\delta,\pm 2\delta,0)$, i.e.\ purely transverse to the main peak [Fig.~\ref{fig:fig1}(a)]. We find $\delta\! =\! 3.0(3)\!\times\! 10^{-3}\;\text{r.l.u.}\! =\! 3.7(3)\!\times\! 10^{-3}\;\text{\AA}^{-1}$ for the $(\delta 00)$-type satellites at all temperatures below $T_N$, corresponding to a modulation with $170(15)$~nm wavelength. For the $(\mp\delta,\pm 2\delta,0)$-type satellites, $\delta$ decreases with temperature, as shown in Fig.~\ref{fig:fig2}(a). No satellite reflections were observed in S4, which we attribute to the poor surface quality of this sample (See Supplemental Material \cite{Note1}). Magnetic correlation lengths extracted from the satellite reflections were found to be nearly isotropic and on the order of 100~nm for all samples \cite{Note1}. No spatial variation was observable when scanning the beam across a given sample surface and we can conclude that magnetic domains must be smaller than the 200~$\mu$m beam dimensions.

The observed symmetry breaking between the satellite reflections at each $\bm{Q}_0$
reveals that these $\bm{Q}_0$ belong to distinct $2Q$ domains, rather than a single-domain multi-$Q$ structure. In particular, the satellite reflections observed at $(\pm\delta,\tfrac{1}{2},0)$ and $(\tfrac{1}{2}\mp\delta,\pm 2\delta,0)$ in S3 correspond to different magnetic wavevectors not related by any symmetry of the paramagnetic phase, so must correspond to different $\bm{Q}$-domains. Such a sample and domain dependence of the satellite wavevectors indicates that the particular long-wavelength magnetic modulation in a given sample is likely selected by a symmetry-breaking field that could be a result of small, local, lattice strains that are quenched in during crystal synthesis, or local correlated defects and stoichiometric variations \cite{goodge2023consequences,mangelsen2021interplay,du2021topological,tanaka2022large}.

Fig.~\ref{fig:fig2} shows the temperature-dependent integrated intensities at $\bm{Q}_0\!=\!(\tfrac{1}{2}00)$ and $\bm{Q}_0\!+\!(\overline{\delta}\,2\delta\,0)$ from S3. Both commensurate and satellite magnetic reflections have a critical temperature of $T_N = 28.5$~K where the Hall response becomes finite [Fig.~\ref{fig:fig2}(b)]. We also observed a smooth decrease in the magnitude of the transverse satellite wave vector as temperature decreases [Fig.~\ref{fig:fig2}(a)], characteristic of a helical magnetic modulation \cite{izyumov1984modulated}. Fixed-$\bm{Q}$ energy scans across the 778.5~eV resonance are typical for Co$^{2+}$ \cite{herrero2015direct,subias2021magnetic} and further confirm the magnetic origin of all observed peaks [inset of Fig.~\ref{fig:fig2} and Fig.~\ref{fig:fig3}(b)].

\subsection{Circular dichroism}\label{subsection:CircDi}

\begin{figure}[t!]
   \includegraphics{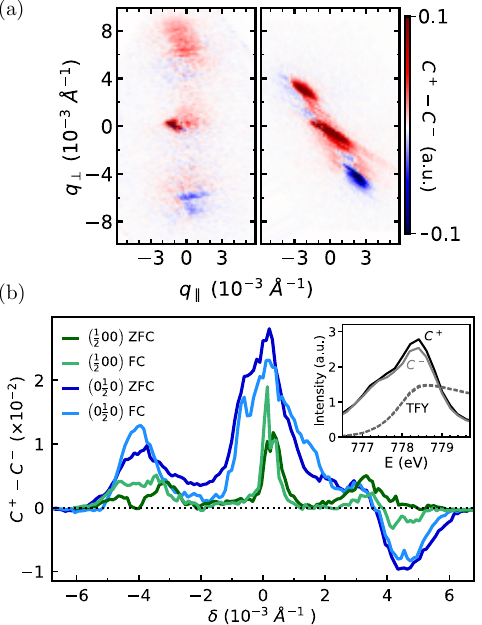}
   \caption{
    (a) Field cooled (FC) CD.
    (b) FC vs ZFC cuts along $(\delta,-2\delta,0)$ at $(\tfrac{1}{2}00)$ and along $(\delta 0 0)$ at $(0\tfrac{1}{2}0)$, averaged over 0.0025~\AA$^{-1}$ along the transverse direction.
    The normalized CD is defined as $(C^+\!-\!C^-)/\int(C^+\!+\!C^-)$. Inset shows energy scans at fixed $\bm{Q} = (\tfrac{1}{2}-\delta,2\delta,0)$.
    }
   \label{fig:fig3}
\end{figure}

Further details of the magnetic structure are revealed by measuring the circular dichroism in the REXS intensity (CD-REXS). CD-REXS is distinct from optical circular dichroism, which is related to the off-diagonal component of the optical conductivity and sensitive to time-reversal symmetry breaking. On the other hand, CD-REXS at a particular wavevector $\bm{Q}$ depends on the orientation of the Fourier component of the spins at this wavevector. 

We performed CD-REXS measurements on S3, where we observe a non-zero CD-REXS signal at all magnetic reflections as shown in Fig.~\ref{fig:fig3}. From our structure factor calculations (see Supplemental Material \cite{Note1}), we find that a finite CD-REXS signal at each peak requires that the ordering associated with each wavevector is noncollinear. As elaborated in the following section~\ref{subsection:FLPA}, the observed circular dichroism indicates that the commensurate component of the magnetic structure is canted out of the basal ($a$-$b$) plane and the incommensurate component is a helix-like structure. Unlike a typical helical spin structure, the CD at the two satellites $\bm{Q}_0\pm\bm{q}$ is not strictly required to have opposite sign \cite{mulders2010circularly, hiraoka2011spinchiral, kim2022origin}. Instead, the total CD at the satellites is offset by the canting angle $\nu$ of the commensurate component \cite{Note1}.

We also observed a variation of the CD along the transverse direction that suggests the presence of domains within the scattering volume. As shown in Fig.~\ref{fig:fig3}(b), we found that the CD varies between subsequent ZFC and 0.1~T FC measurements, particularly for the $(\tfrac{1}{2}00)$ peaks. This suggests that these domains may be related by time-reversal, similar to the domains observed in optical CD measurements \cite{gu2025probing}.


\begin{figure}[h!]
    \includegraphics{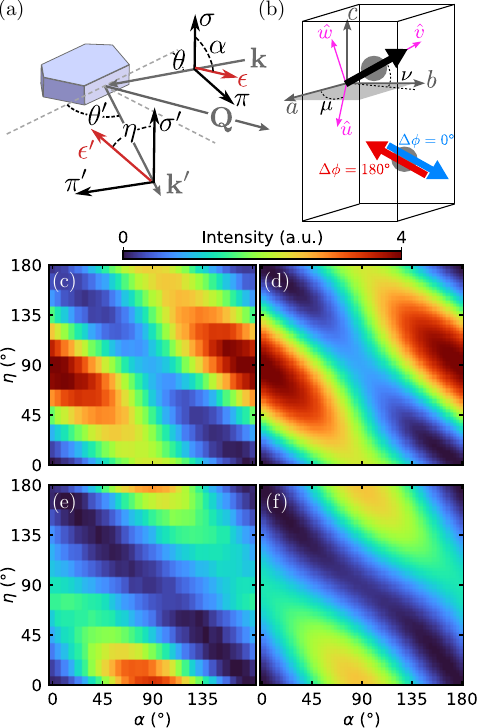}
    \caption{
    (a) Experimental geometry for REXS measurements.
    (b) Fourier components $\bm{S}_{\bm{Q}_0,1}$ and $\bm{S}_{\bm{Q}_0,2}$ of the commensurate part of the magnetic structure, showing the in-plane angle $\mu$ and canting angle $\nu$ for the relative phases
    $\Delta\phi = 0\degree$ (blue) and
    $\Delta\phi = 180\degree$ (red).
    (c) measured FLPA at $(\frac{1}{2}00)$ 
    (d) calculated FLPA at $(\frac{1}{2}00)$
    (e) measured FLPA at $(0\frac{1}{2}0)$
    (f) calculated FLPA at $(0\frac{1}{2}0)$
    }
    \label{fig:flpa}
\end{figure}

\begin{figure*}[t!]
    \includegraphics{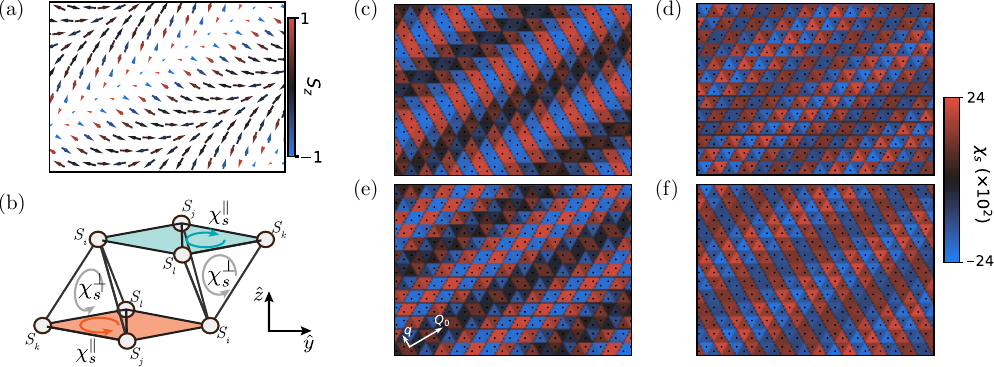}
    \caption{
    (a) Real space depiction of magnetic structure on a single sublattice for $\bm{Q}_0\! =\! (\tfrac{1}{2}00)$ and $\bm{q}\! =\! (-\delta,2\delta,0)$ with $\delta \!=\! 0.053$ r.l.u., $\phi \!=\! 45\degree$, $\angIP{} \!=\! 110\degree$, $\angOP{} \!=\! 30\degree$. (b) Co$^{2+}$ triangular plaquettes used to define the intra- and inter-sublattice scalar spin chirality contributions $\chi_s^{\parallel}$ and $\chi_s^{\perp}$.
    Scalar spin chirality contribution from the inter-sublattice plaquettes $\chi_s^{\perp}$ for the spin configuration in (a) with 
    (c) $\Delta\phi = 0\degree$, $\Delta\psi\!=\!0\degree$,
    (d) $\Delta\phi = 0\degree$, $\Delta\psi\!=\!180\degree$,
    (e) $\Delta\phi = 180\degree$, $\Delta\psi\!=\!0\degree$,
    (f) $\Delta\phi = 180\degree$, $\Delta\psi\!=\!180\degree$,
    The color of each triangle represents $\chi_s$ projected onto the $z$ component of the plaquette and summed over the three plaquettes sharing a vertex plotted as black dots.
    }
    \label{fig:ssc}
\end{figure*}


\subsection{Full linear polarization analysis}\label{subsection:FLPA}

Orientations of the magnetic Fourier components $\bm{S}_{\bm{Q}_0,n}^{}$ were determined from full linear polarization analysis (FLPA) of the REXS intensity\cite{wang2012complete,mazzoli2007disentangling,johnson2008determination,frawley2017elucidation,herreromartin2015direct,subias2021magnetic,rahn2018coupling,simeth2023rexs} by measuring the intensity at $\bm{Q}_0$ as a function of incident polarization angle $\alpha$ and polarization analyzer angle $\eta$ [Fig.~\ref{fig:fig1}(a)]. We consider a $2Q$ magnetic structure with propagation vectors $\bm{Q}_0$ and $\bm{Q}_0\pm\bm{q}$, similar to the one considered in \cite{heinonen2022magnetic}: 
\begin{eqnarray}\label{eq:spin_structure}
    \bm{S}_n^{}(\bm{r}_j)
    &=& S \sin\phi_n \cos(\bm{Q}_0 \cdot \bm{r}_j) \hat{\bm{v}}_n \nonumber \\
    &+& S \cos\phi_n \cos[(\bm{Q}_0+\bm{q}) \cdot \bm{r}_j+\psi_n] \hat{\bm{u}}_n \nonumber \\
    &+& \chi S \cos\phi_n \sin[(\bm{Q}_0+\bm{q}) \cdot \bm{r}_j+\psi_n] \hat{\bm{w}}_n,
\end{eqnarray}
where $n\!=\!1,2$ labels the sublattices at each of the 2 Co sites in the unit cell, $\chi\! =\! \pm 1$ is the helix chirality, $\phi_n$ and $\psi_n$ are the phases on sublattice $n$ for $\bm{Q}_0$ and $\bm{Q}_0\! \pm\! \bm{q}$ respectively, and $\hat{\bm{u}}_n$, $\hat{\bm{v}}_n$, and $\hat{\bm{w}}_n$ are unit vectors, assumed to be orthogonal to maintain a constant moment size. We note that this structure contains two magnetic propagation vectors of unequal magnitude, and so is distinct from the typical $2Q$ structures, which consist of two symmetry-equivalent wavevectors \cite{marmorini2014magnon,ozawa2016vortex,hayami2021topological}. The Fourier components of this structure are
\begin{eqnarray}
    \bm{S}_{\bm{Q}_0,n}^{} &=& \sqrt{N}S\sin\phi_n \hat{\bm{v}}_n, \nonumber \\
    \bm{S}_{\bm{Q}_0\pm\bm{q},n}^{} &=& \sqrt{N}\frac{S}{2}\cos\phi_n e^{i\psi_n} (\hat{\bm{u}}_n \pm i\chi\hat{\bm{w}}_n),
\end{eqnarray}
where $\bm{S}_{\bm{Q}_0,n}^{} = \bm{S}_{-\bm{Q}_0,n}^* = \bm{S}_{\bm{Q}_0,n}^*$, $\bm{S}_{\bm{Q}_0\pm\bm{q},n}^{} = \bm{S}_{-\bm{Q}_0\mp\bm{q},n}^*$, and $N$ is the number of unit cells. We parameterize the magnetic structure with the angle $\angIP{n}$ between $\hat{\bm{u}}_n$ and the lattice vector $\bm{a}_1\! =\! a\hat{\bm{x}}$ in the $a$-$b$ plane, out-of-plane canting angle $\angOP{n}$ of $\hat{\bm{v}}_n$, and phases $\phi_n$ and $\psi_n$. The angles $\angIP{}$ and $\angOP{}$ are defined in Fig.~\ref{fig:flpa}(b). We fit the commensurate component $\bm{S}_{\bm{Q}_0,n}^{}$ to the measured FLPA shown in Fig.~\ref{fig:flpa}(c) and (e). $\angOP{1}\!=\!\angOP{2}$ is ruled out as this always leads to zero $\pi$-$\pi'$ intensity, inconsistent with the observed FLPA and CD. For our analysis, we have fixed $\angIP{1}\!=\!\angIP{2}\!=\!\angIP{}$ and $\angOP{1}\!=\!-\angOP{2}\!=\!\angOP{}$. We find no improvements by relaxing these constraints. 

The phases $\phi_n$ determine the relative amplitude of the commensurate and incommensurate component for each sublattice while the phases $\psi_n$ define an overall translation of the incommensurate modulation. The specific values of the $\psi_n$ are arbitrary, but the relative phase $\Delta\psi = \psi_2 - \psi_1$ is relevant. Symmetry constrains $\Delta\phi\!=\!\phi_2\!-\!\phi_1$ to either $0\degree$ or $180\degree$ \cite{wu2022highly}. The specific value of $\phi = \phi_1 = \phi_2 + \Delta\phi$ only determines the relative intensities of the main and satellite peaks; but, the intensity of the main peak is independent of $\Delta\psi$. Thus, we can model the FLPA data using only the parameters $\angIP{}$ and $\angOP{}$, for the two cases of $\Delta\phi\!=\!0\degree$ or $\Delta\phi\!=\!180\degree$. The relative orientations of the commensurate Fourier components $\bm{S}_{\bm{Q}_0,1}$ and $\bm{S}_{\bm{Q}_0,2}$ for these two cases are shown in Fig.~\ref{fig:flpa}(b).

We find $\angIP{}\! =\! 109(1)\degree$ at $(\tfrac{1}{2}00)$ and $\angIP{}\! =\! 12(1)\degree$ at $(0\tfrac{1}{2}0)$ for each case, or nearly $\pm 80\degree$ from $\bm{Q}_0$. The in-plane angle relative to $\bm{Q}_0$ is opposite in each domain, with the same broken symmetry as the modulation wavevectors in S1 and S2 \cite{Note1}. For $\Delta\phi=0\degree$ we find $\angOP{} = 37(2)\degree$ at $(\tfrac{1}{2}00)$ and $\angOP{} \!=\! 24(2)\degree$ at $(0\tfrac{1}{2}0)$. While for $\Delta\phi \! = \! 180\degree$, we find $\angOP{} \! = \! 14(2)\degree$ at $(\tfrac{1}{2}00)$ and $\angOP{} \! = \! 9(2)\degree$ at $(0\tfrac{1}{2}0)$. Both cases adequately describe the data at $(\tfrac{1}{2}00)$ while neither fully matches the intensity at $\pi$-$\sigma'$ for $(0\tfrac{1}{2}0)$. We attribute this discrepancy to a slight analyzer misalignment. Furthermore, we cannot rule out contributions from domains with different moment orientations. The results of our fit are summarized in Table~\ref{tab:flpa_parameters} and Fig.~\ref{fig:ssc}(a) shows a real-space representation of the non-coplanar magnetic structure found in S3.

After determining $\angIP{}$ and $\angOP{}$, we can estimate the value of $\phi = \phi_1$ and $\phi_2 = \phi_1+\Delta\phi$ from the relative integrated intensities of the main and satellite peaks. Using the measured ratio $I(\bm{Q}_0)/I(\bm{Q}_0\pm\bm{q}) \approx 1.5$ and the computed structure factor (see Supplemental Material \cite{Note1}), we find $\phi$ between 20$\degree$ and 45$\degree$. We note that our measurements do not probe the weak out-of-plane ferromagnetic component at $Q=0$, known to be connected to the AHE \cite{tanaka2022large}. 


We also compare our results with the proposed tetrahedral $3Q$ structure for \cns \cite{takagi2023spontaneous,park2023tetrahedral,dong2024simple}. While our measurements are consistent with the polarized neutron diffraction measurements that constrain the magnetic structure to have out-of-plane components, 
the tetrahedral $3Q$ structure has Fourier components that are collinear between each sublattice for a given $\bm{Q}$, forbidding any non-zero CD and $\pi$-$\pi'$ intensity. Thus, our CD-REXS and FLPA results rule out the tetrahedral $3Q$ structure for our \cns samples that display large AHE. We have also computed the FLPA pattern for the tetrahedral $3Q$ structure and find that it is inconsistent with the data (See Supplemental Material \cite{Note1}).

\renewcommand{\arraystretch}{1.5}
\begin{table}[t!]
  \centering
  \caption{Parameters obtained from FLPA describing the commensurate Fourier component, for the two possible choices of $\Delta\phi$.}
  \begin{tabular}{c c c c c}
  \hline\hline
  {}                 & {}    &   $\Delta\phi=0\degree$ & $\Delta\phi=180\degree$ \\ \hline
  Peak               & $\mu$ & $\nu$  & $\nu$ \\ \hline
  $(\tfrac{1}{2}00)$ & 109(1)$\degree$ & 37(2)$\degree$ & 14(2)$\degree$ \\
  $(0\tfrac{1}{2}0)$ & 12(1)$\degree$  & 24(2)$\degree$ & 9(2)$\degree$ \\
  \hline\hline
  \end{tabular}
  \label{tab:flpa_parameters}
\end{table}

\subsection{Scalar spin chirality}\label{subsection:ScalarChi}

In the absence of a uniform magnetization, a Hall response can be generated in non-coplanar antiferromagnets through a uniform scalar spin chirality $\chi_s^{}\! =\! \bm{S}(\bm{r}_i)\cdot[\bm{S}(\bm{r}_j)\times\bm{S}(\bm{r}_k)]$ with sites $i$, $j$, $k$ on a triangular plaquette. Conduction electrons traversing closed loops around regions of uniform $\chi_s$ accumulate a net Berry phase \cite{nagaosa2010anomalous}, while a Berry phase can accumulate for a non-uniform $\chi_s$ when spin-orbit coupling is present \cite{zhang2020real}. To check these possibilities in \cns{}, we compute $\chi_s$ for the $2Q$ chiral stripe magnetic structure. Since there are two triangular lattice layers per unit cell in \cns{}, there may be contributions to the total spin chirality from both intra-sublattice plaquettes within the triangular lattice basal plane $\chi_s^{\parallel}$, and inter-sublattice plaquettes $\chi_s^{\perp}$ that involve two sites from one sublattice and one site from the opposite one. We compute the total scalar spin chirality using the real space spin structures found above by considering separate contributions $\chi_s^{\parallel}$ and $\chi_s^{\perp}$ [Fig.~\ref{fig:ssc}(b)]. We find that a finite incommensurate modulation $q \neq 0$ gives rise to a staggered scalar chirality in \cns{}, with a specific form that depends on the relative phase and canting direction $\nu$. For all physical choices of $\nu$ and relative phase parameters \cns develops a staggered striped or checkerboard pattern of scalar spin chirality modulated along both $\bm{Q}_0$ and $\bm{q}$ as shown in Fig.~\ref{fig:ssc}(c-f). Such a staggered scalar chirality is distinct from the uniform chirality of the tetrahedral triple-$\bm{Q}$ state that was observed in \cts \cite{takagi2023spontaneous,park2023tetrahedral} and is not expected to directly generate a Hall response via the real-space Berry curvature. However, as we will explain in section~\ref{section:AHE}, the $2Q$ chiral stripe ordering we observe in S3 does in fact break all symmetries of the crystal structure that would otherwise forbid a Hall response, and thus is essential for generating that response. We address below how such a complex $2Q$ scalar chiral stripe ordering can naturally arise from four-spin interactions in this metallic magnet.


\section{Origin of Satellite Magnetic Bragg Peaks} \label{section:small-q}

In this section, we explain the origin of the observed satellite magnetic Bragg peaks from four-spin interactions in \cns. The physical mechanism is most conveniently illustrated by considering a single-layer triangular lattice antiferromagnetic model, which includes isotropic Heisenberg and four-spin interactions:
\begin{equation} \label{eq:hamil}
    \mathcal{H} = \sum_{\bm{k}} J_{\bm{k}}\bm{S}_{\bm{k}} \cdot \bm{S}_{-\bm{k}} 
    + \frac{1}{N} \sum_{\bm{p},\bm{k},\bm{l}} K_{\bm{p}\bm{k}\bm{l}}
    (\bm{S}_{\bm{p}} \cdot \bm{S}_{\bm{k}}) 
    (\bm{S}_{\bm{l}} \cdot \bm{S}_{-\bm{k}-\bm{p}-\bm{l}}),
\end{equation}
where $N$ is the total number of spins, $J_{\bm{k}} = J_{-\bm{k}}$ for a symmetric spin-exchange, the vertex $K_{\bm{p}\bm{k}\bm{l}}$ is symmetrized with respect to the exchange of spins in each scalar product and the exchange of the two scalar products, and $\bm{S}_{\bm{k}}$ is the Fourier transform of the real space spin variables ${\cal\bm S}_{\bm{r}}$,
\begin{equation}
    \bm{S}_{\bm{k}} = \frac{1}{\sqrt{N}} \sum_{\bm{r}} e^{i\bm{k}\cdot\bm{r}} \mathcal{\bm{S}}_{\bm{r}}.
\end{equation}
To understand the origin of the magnetic ordering, it is enough to consider the classical limit, meaning that we will replace the spin operators with vectors of norm $S$:
\begin{equation} \label{eq:rsc}
    \mathcal{\bm{S}}_{\bm{r}} \cdot \mathcal{\bm{S}}_{\bm{r}} = S^2.
\end{equation}
This real space constraint leads to the following constraint in momentum space:
\begin{equation} \label{eq:gc}
    \sum_{\bm{k}} \bm{S}_{\bm{k}} \cdot \bm{S}_{-\bm{k}} = NS^2.
\end{equation}

In absence of the four-spin interaction, the ground state of the classical Hamiltonian ${\mathcal{H}}$ is any spiral ordering with wave vector $\bm{Q}_0$ that minimizes the exchange interaction $J_{\bm{k}}$. This means that 
\begin{equation}
    \bm{S}_{\bm{k}} = \bm{S}_{\bm{Q}_0} \delta_{\bm{k},\bm{Q}_0}.
\end{equation}
Note that we are considering a single layer, and the sublattice index $n$ of the Fourier component has been dropped. For the case of interest, $\bm{Q}_0 = -\bm{Q}_0$ is any of the three M points of the Brillouin zone of the triangular lattice.  
For this particular case, the $1Q$ ordering has exactly the same energy as any multi-$\bm{Q}$ ordering of the form:
\begin{equation}
    \bm{S}_{\bm{k}} = \bm{S}_{\bm{Q}_0} \delta_{\bm{k},\bm{Q}_0} + \bm{S}_{\bm{Q}_1}\delta_{\bm{k},\bm{Q}_1} +
    \bm{S}_{\bm{Q}_2}\delta_{\bm{k},\bm{Q}_2}
\end{equation}
where $\bm{Q}_1$ and $\bm{Q}_2$ are the wave vectors associated with the other M-points of the Brillouin zone,
while the three vector amplitudes $\bm{S}_{\bm{Q}_{\nu}}$ ($\nu = 0,1,2$) are mutually orthogonal and obey the normalization condition \eqref{eq:gc} $|\bm{S}_{\bm{Q}_0}|^2 + |\bm{S}_{\bm{Q}_1}|^2 + |\bm{S}_{\bm{Q}_2}|^2 = NS^2$. This continuous degeneracy of isotropic Heisenberg models can either be removed by anisotropic interactions or by the isotropic four-spin interactions included in ${\mathcal{H}}$. For instance four-spin interactions 
$K_{\bm{Q}_{\nu} \bm{Q}_{\nu} \bm{Q}_{\nu}}<0$ and $K_{\bm{Q}_{\nu} \bm{Q}_{\nu} \bm{Q}_{\nu'}}>0$ with $\nu \neq \nu'$ favor the $1Q$ ordering that is relevant for our analysis.

Our next step is to demonstrate that incommensurate Fourier satellite peaks around the dominant commensurate peak at $\bm{Q}_0$ can be induced by a finite four-spin interaction $K_{\bm{p}\bm{k}\bm{l}}$. For this purpose, we consider the following ansatz
\begin{equation} \label{eq:pert}
    \bm{S}_{\bm{k}} = 
    \bm{S}_{\bm{Q}_0}\delta_{\bm{k},\bm{Q}_0} + 
    \bm{S}_{\bm{Q}_0+\bm{q}}\delta_{\bm{k},\bm{Q}_0+\bm{q}} + 
    \bm{S}^*_{\bm{Q}_0+\bm{q}}\delta_{\bm{k},-\bm{Q}_0-\bm{q}},
\end{equation}
where the amplitudes $\bm{S}_{\bm{Q}_0+\bm{q}} = \bm{S}^*_{\bm{Q}_0-\bm{q}}$ represents the ``satellite components'' on top of the dominant $1Q$ ordering, while $\bm{S}_{\bm{Q}_0} = \bm{S}_{\bm{Q}_0}^*$ represents the dominant collinear component. The global constraint \eqref{eq:gc} implies that:
\begin{equation}
    |\bm{S}_{\bm{Q}_0}|^2 + 2|\bm{S}_{\bm{Q}_0+\bm{q}}|^2 = NS^2.
\label{eq:const2}
\end{equation}
Its Fourier transform gives rise to the real-space spin configuration
\begin{eqnarray} \label{eq:rspace}
    \bm{S}(\bm{r}) &=& \frac{1}{\sqrt{N}}\left[
    \bm{S}_{\bm{Q}_0} \cos(\bm{Q}_0\cdot\bm{r}) + \bm{S}_{\bm{Q}_0+\bm{q}} e^{i (\bm{Q}_0+\bm{q}) \cdot \bm{r}} \right. \nonumber \\
    && \left. + \bm{S}^*_{\bm{Q}_0+\bm{q}} e^{-i(\bm{Q}_0+\bm{q})\cdot\bm{r}} \right].
\end{eqnarray}
The real-space constraint in Eq.~(\ref{eq:rsc}) implies that $\bm{S}_{\bm{Q}_0+\bm{q}} \cdot \bm{S}_{\bm{Q}_0} = \bm{S}_{\bm{Q}_0 + \bm{q}}^* \cdot \bm{S}_{\bm{Q}_0} = 0$ and $\bm{S}_{\bm{Q}_0+\bm{q}} \cdot \bm{S}_{\bm{Q}_0+\bm{q}} = 0$. Note that the last condition is feasible because this quantity $\bm{S}_{\bm{Q}_0+\bm{q}} \cdot \bm{S}_{\bm{Q}_0+\bm{q}}$ is not positive defined. Under these conditions together with Eqs.~\eqref{eq:rspace} and \eqref{eq:const2}, the real-space constraint in Eq.~(\ref{eq:rsc}) is explicitly satisfied. The above analysis implies that one can parameterize the spin configuration using three mutually orthogonal unit vectors ($\hat{\bm{w}},\hat{\bm{u}},\hat{\bm{v}}$):
\begin{eqnarray}
    \bm{S}_{\bm{Q}_0} &=& \sqrt{N}S\sin\phi\hat{\bm{v}}, \nonumber \\
    \bm{S}_{\bm{Q}_0+\bm{q}} &=& \sqrt{N}{S\over 2}\cos\phi e^{i\psi}(\hat{\bm{u}}- i\chi\hat{\bm{w}}),
\end{eqnarray}
where $\chi=\pm1$. This momentum space spin structure corresponds to the real-space configuration
\begin{eqnarray}\label{eq:RS_SpinConfig}
    \bm{S}(\bm{r}) &=&  S\sin{\phi}\cos(\bm{Q}_0\cdot\bm{r}) \hat{\bm{v}} \nonumber \\
    &+& S \cos{\phi}\cos[(\bm{Q}_0+\bm{q})\cdot \bm{r}+\psi] \hat{\bm{u}} \nonumber \\
     &+& \chi S\cos{\phi}\cos[(\bm{Q}_0+\bm{q})\cdot \bm{r}+\psi] \hat{\bm{w}},
\end{eqnarray}
consistent with the measured one (Eq. \ref{eq:spin_structure}). The unit vector $\hat{\bm{v}}$ denotes the spin direction of the dominant collinear component, while the unit vectors $\hat{\bm{u}}$ and $\hat{\bm{w}}$ span the polarization plane of the $\bm{q}$-spiral. The angle $\phi$ determines the relative intensity between the dominant and satellite peaks, and the phase $\psi$ gives rise to degenerate ground states related by translation.


The optimal magnitude of $\bm{S}_{\bm{Q}_0+\bm{q}}$ can be obtained by inserting expression \eqref{eq:pert} into Eq.~\eqref{eq:hamil} and keeping the leading order non-trivial contribution, which yields
\begin{eqnarray} \label{eq:hamil2}
    E &\simeq & N  (J_{\bm{Q}_0}S^2 + S^4K_{\bm{Q}_0 \bm{Q}_0 \bm{Q}_0}) + 2\left[J_{\bm{Q}_0+\bm{q}} - J_{\bm{Q}_0} \right. \nonumber \\  
     && \left. + 2S^2(K_{\bm{Q}_0,\bm{Q}_0,\bm{Q}_0+\bm{q}} - K_{\bm{Q}_0\bm{Q}_0\bm{Q}_0}) \right] |\bm{S}_{\bm{Q}_0+\bm{q}}|^2  
     \nonumber \\
     &+& \frac{1}{N} \left(4K_{\bm{Q}_0+\bm{q},\bar{\bm{Q}}_0+\bar{\bm{q}},\bm{Q}_0+\bm{q}} - 8K_{\bm{Q}_0,\bm{Q}_0,\bm{Q}_0+\bm{q}} \right. \nonumber \\
     && \left. + 4K_{\bm{Q}_0\bm{Q}_0\bm{Q}_0} \right) |\bm{S}_{\bm{Q}_0+\bm{q}}|^4
\end{eqnarray}
where $\bar{\bm{q}} \equiv -\bm{q}$. As previously mentioned, without the four-spin interaction, the quadratic term is positive definite, which is a condition for M-point ordering, and the system would retain the collinear $1Q$ order. However, in the presence of a four-spin term, the quadratic term can become negatively defined along one of its principal axes, converting the M-point into a saddle point. In this scenario, a finite magnitude of $\bm{S}_{\bm{Q}_0+\bm{q}}$ develops to lower the ground state energy. Given the two-fold rotation symmetries in the little group of the M-point, the principle axes must point along the $\bm{Q}_0$ and $\bm{Q}_0 \times \hat{\bm{z}}$ directions. We then align the coordinate system in momentum space along the principle axes of the quadratic term and expand the energy function for small $\bm{q}$:
\begin{eqnarray} \label{eq:hamil3}
    E &\simeq& N(J_{\bm{Q}_0}S^2 + S^4K_{\bm{Q}_0\bm{Q}_0\bm{Q}_0}) \nonumber \\
    &+& \tilde{q}_{\mu}^2
    \left[\frac{\partial^2 J_{\bm{k}}}{\partial \tilde{k}_{\mu}^2} + 2S^2 
    \frac{\partial^2 K_{\bm{Q}_0\bm{Q}_0\bm{k}}}{\partial \tilde{k}_{\mu}^2} \right]_{\bm{k}=\bm{Q}_0} |\bm{S}_{\bm{Q}_0+\bm{q}}|^2 
    \nonumber \\
    &+& \frac{2\tilde{q}_{\mu}^2}{N}
    \left[ \frac{\partial^2 K_{\bm{k}\bar{\bm{k}}\bm{k}}}{\partial \tilde{k}_{\mu}^2} - 2\frac{\partial^2 K_{\bm{Q}_0\bm{Q}_0\bm{k}}}{\partial \tilde{k}_{\mu}^2} \right]_{\bm{k}=\bm{Q}_0} |\bm{S}_{\bm{Q}_0+\bm q}|^4,
    \nonumber \\
\end{eqnarray}
where we are adopting the convention of summation over repeated Greek letters. Assuming that there is a principal direction $\mu_0$ for which,
\begin{eqnarray}\label{eq:satelliteCondition}
    \left[ \frac{\partial^2 J_{\bm{k}}}{\partial \tilde{k}_{\mu_0}^2} + 2S^2 \frac{\partial^2 K_{\bm{Q}_0\bm{Q}_0\bm{k}}}{\partial \tilde{k}_{\mu_0}^2} \right]_{\bm{k}=\bm{Q}_0} &<& 0,
    \nonumber \\
    \left[ \frac{\partial^2 K_{\bm{k}\bar{\bm{k}}\bm{k}}}{\partial \tilde{k}_{\mu_0}^2} - 2\frac{\partial^2 K_{\bm{Q}_0\bm{Q}_0\bm{k}}}{\partial \tilde{k}_{\mu_0}^2} \right]_{\bm{k}=\bm{Q}_0} &>& 0,
\end{eqnarray}
a finite satellite component is obtained with intensity:
\begin{equation}
    \frac{|\bm{S}_{\bm{Q}_0+\bm{q}}|^2}{N} = 
    -\frac{1}{4}\left[
    \frac{\frac{\partial^2}{\partial\tilde{k}_{\mu_0}^2}
    \left(J_{\bm{k}} + 2S^2K_{\bm{Q}_0\bm{Q}_0\bm{k}} \right)}{ \frac{\partial^2}{\partial \tilde{k}_{\mu_0}^2}
    \left(K_{\bm{k}\bar{\bm{k}}\bm{k}} - 2K_{\bm{Q}_0\bm{Q}_0\bm{k}} \right)}
    \right]_{\bm{k}=\bm{Q}_0},
\end{equation}
where $\bm{q}$ is parallel to either the  $\bm{Q}_0$ or $\bm{Q}_0 \times \hat{\bm{z}}$ directions. The magnitude of optimal $\bm{q}$ is obtained by extending the above Taylor expansion in $\bm{q}$ up to quartic order, which yields
\begin{eqnarray}
    \rvert \bm{q} \rvert = -6\left[ 
    \frac{\frac{\partial^2}{\partial\tilde{k}_{\mu_0}^2} \left(J_{\bm{k}} + 2S^2K_{\bm{Q}_0\bm{Q}_0\bm{k}} \right)}{\frac{\partial^4}{\partial\tilde{k}_{\mu_0}^4} \left(J_{\bm{k}} + 2S^2K_{\bm{Q}_0\bm{Q}_0\bm{k}} \right)}
    \right]_{\bm{k}=\bm{Q}_0}.
\end{eqnarray}

We conclude from the above analysis that the four-spin interaction can induce the satellite peaks observed from experiment. Importantly, this mechanism only relies on the second derivatives of the bilinear and four-spin interaction around $\bm{Q}_0$, leaving $J_{\bm{Q}_0}$ and $K_{\bm{Q}_0\bm{Q}_0\bm{Q}_0}$ as free parameters that stabilize the dominant $1Q$ ordering. The direction of $\bm{q}$ would be along or perpendicular to $\bm{Q}_0$ according to the above simple analysis, which is consistent with the observation in S3. The observation of satellites along non-transverse directions indicates the possibility of some subtle structural symmetry-breaking effects, discussed below in section~\ref{subsection:slanted&domains}.

\section{Symmetry breaking and Anomalous Hall Effect} \label{section:AHE}

Having identified that the magnetic order in \cns is a $2Q$ chiral stripe phase with staggered scalar spin chirality, we now address whether such a magnetic ordering is consistent with the anomalous Hall effect (AHE). We consider two separate cases corresponding to the distinct types of satellite peaks observed. Unless otherwise stated, all satellite wavevectors below are relative to the commensurate peak at $\bm{Q}_0 = (\tfrac{1}{2}00)$.

The first case corresponds to transverse satellite peaks $\bm{q}_{\text{M}}^{} = (\mp \delta, \pm 2\delta, 0)$ as in S3. The second case corresponds to ``slanted'' satellites $\bm{q}_{\text{M}'}^{} = (0,\pm \delta,0)$ and $\bm{q}_{\text{M}''}^{} = (\pm \delta,0,0)$, as in S1 and S2 [Fig.~\ref{fig:domains_symmetry}(c) and (d)]. For each case, we determine whether the magnetic order breaks crystalline symmetries precluding the AHE, as described in Section~\ref{subsection:precluding}. In addition to addressing the AHE, we also reconcile the coexistence of the two cases on the basis of symmetry.

In Section~\ref{subsection:transverse}, we will see that in domains where transverse satellites are present, the residual symmetry group of the magnetic order is compatible with the AHE or uniform magnetization along the $c$-axis (weak ferromagnetism). When slanted satellites are present, the residual symmetry of the magnetic order includes the composition of time-reversal and a lattice translation perpendicular to $\bm{q}$, thus precluding the AHE, as explained in Section~\ref{subsection:slanted&domains}. As the AHE is measured across all samples S1-S3, we propose a possible explanation from structural symmetry lowering.

The observations of both the transverse and slanted satellites, not related to one another by symmetry, indicates the possibility that the symmetry of the \cns structure has been broken. For this reason, we outline in Section~\ref{subsection:slanted&domains} the possible residual structural symmetry groups, identifying those which are best able to justify the measurements across samples S1-S3.

\subsection{Symmetries precluding the AHE in \texorpdfstring{\cns}{}} \label{subsection:precluding}

The crystal lattice of \cns is reported to be in the chiral space group $P6_{3}22$ (No. 182) \cite{anzenhofer1970crystal,parkin1983magnetic}. Among its point elements, this group contains a six-fold screw axis along the $c$-direction, $[C_{6z}|00\frac{1}{2}]$, and two-fold axes perpendicular to $c$, such as $[C_{2x}|000]$ and its composition with powers of the six-fold element. These two-fold axes pass through the corners and edge-centers of the hexagons formed by the Nb atoms. Together with time-reversal symmetry $\tau$ in the paramagnetic phase, these symmetries give rise to the gray magnetic space group $P6_{3}22.1'$. Assuming that the transition is second order, consistent with the experiments [Fig.~\ref{fig:fig2}]  the symmetry will be reduced to a subgroup of $P6_{3}22.1'$ upon cooling below $T_N$. We first restrict our attention to the point subgroups of the gray point group $622.1'$ of the paramagnetic phase. The gray point group $622.1'$ is generated by the symmetry elements explicitly listed above (modulo lattice translations).

Of the nineteen subgroups of $622.1'$, eight allow for the presence of an AHE ($62'2'$, $6$, $32'$, $3$, $2'2'2$, $2'$, $2$, and the trivial group $1$), indicated by the non-zero symmetry-allowed form of the antisymmetric part of the AHC tensor for each of (magnetic) point groups \cite{mtensorBilbaoAHCtensors}. The common feature of these subgroups is the breaking of $\tau$, $\tau C_{6z}$, and $C_{2x}$ symmetries. We can therefore interpret these three symmetry elements as those that prohibit an AHE. 

In addition to the above point symmetries, the gray space group $P6_{3}22.1'$ contains compositions of time-reversal and translations. All such elements prohibit the anomalous Hall effect, as they change the sign of the Hall conductivity $\sigma_{xy}$. Thus, \emph{any} magnetic order in \cns that breaks these three point symmetries and the composition of time-reversal and translations, will permit the AHE.

\subsection{Transverse satellite wavevectors} \label{subsection:transverse}

In domains with transverse satellites, all symmetries of \cns precluding the AHE are broken. The magnetic order breaks all point symmetries. From Eq.~\ref{eq:RS_SpinConfig} it may also be shown that no translation brings the order to one in which all spins are reversed: this implies all elements composing time-reversal and translation translation are broken. If we consider only a single triangular sublattice with isotropic spins, a transformation composing time-reversal, translation by one lattice space $\delta\bm{r}$ with $\delta\bm{r} \cdot \bm{Q}_0 = \pi$, and a global spin rotation about the $\hat{\bm{v}}$-axis by an angle $\pi-\delta\bm{r}\cdot\bm{q}$ is present. As discussed in Section~\ref{subsection:precluding}, such an element would forbid the AHE. However, when both triangular sublattices are taken into account there is no such symmetry because of the noncollinearity of the commensurate components on each sublattice (Fig.~\ref{fig:flpa}), i.e.\ there are two different $\hat{\bm{v}}$ axes.

Importantly, the transverse satellites are essential for generating a non-zero anomalous Hall response. A $1Q$ magnetic order with propagation vector at an M-point results in a spin configuration where opposite-spin sublattices are related by a lattice translation. Consequently, the combined operation of time reversal and translation remains a symmetry of the system. Thus, the breaking of the symmetries precluding the AHE requires both the incommensurate spiral and the noncollinearity of the commensurate component.


\subsection{Slanted satellites and structural symmetry breaking }\label{subsection:slanted&domains}

\begin{figure*}[ht!]
    \centering
    \includegraphics[scale=1.0]{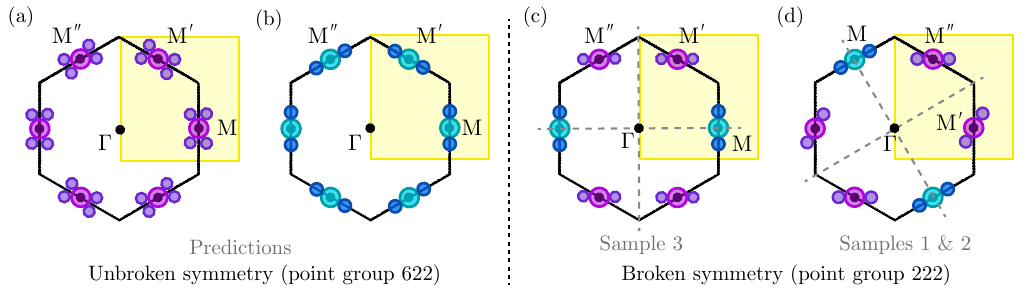}
    \caption{Satellite peaks expected in the cases of (a) all slanted and (b) all transverse if the full symmetry of \cns is retained in the crystal structure. The yellow box indicates the subset of peaks accessible from a given sample surface in reflection geometry.
    Measured (yellow box) and inferred satellite peaks in (c) S3 and (d) S1 and S2. The inferred peaks are those consistent with the lower structural symmetry point group $222$, whose in-plane axes are shown as dashed gray lines.
    }
    \label{fig:domains_symmetry}
\end{figure*}

Domains with slanted satellites $\bm{q} = (0,\pm\delta,0)$ and $\bm{q} = (\pm\delta,0,0)$ as observed in S1, S2, and S3 have the property that translating by one lattice spacing along the $a$ and $b$ axes respectively, reverses the orientation of all spins, as can be seen from Eq.~\ref{eq:RS_SpinConfig}. Thus, the residual symmetry in these domains includes the composition of time-reversal and translations, and the AHE should vanish. However, the presence of the slanted satellites is inconsistent with the theory of Section~\ref{section:small-q}. Furthermore, the sets of magnetic reflections observed around the $(\tfrac{1}{2}00)$ and $(0\tfrac{1}{2}0)$ reciprocal lattice positions are not consistent with domains that are related by the symmetries of the paramagnetic space group $P6_{3}22.1'$. Together, these observations suggest the presence 
of an additional, and not yet detected, symmetry-breaking in \cns that was not considered above. In this section, we explore speculative explanations for the presence of these satellites and the measurement of non-zero AHE in samples that contain only slanted satellites. 

As described in Section~\ref{subsection:precluding}, the full structural point group symmetry of \cns{} is 622. At the onset of magnetic order, this symmetry is reduced, leading to the formation of domains related by each of the broken symmetries. Consider a magnetic order with peaks centered at an M-point. If the crystal maintains the full 622 point symmetry, one expects a symmetry-related domain with peaks centered at $\bm{Q}_{\text{M}'} = C_{6z}\bm{Q}_{\text{M}}$, obtained from the sixfold rotation that is broken by the magnetic order. Consequently, the satellite peaks $\bm{q}_{\text{M}'}$ at M$'$ must also be related to those at M by a $C_{6z}$ rotation. It follows that, from a measurement at a single M-point, we can predict the locations of satellite peaks at all other M-points connected by the symmetries of \cns, as illustrated in Fig.~\ref{fig:domains_symmetry}(a,b). For transverse peaks at M, we expect transverse peaks at all other symmetry-related points. For slanted peaks, we expect two types of $q$-domains at each $\bm{Q}_0$, corresponding to a $\pm 60\degree$ rotation from $\bm{Q}_0$. However, neither of these predictions match the peaks observed in any samples, indicating that there must be a lower structural symmetry.

The measured satellite peaks are consistent with the crystal structure breaking all three-fold and six-fold axes, leaving only the possibility of two-fold axes along the original crystallographic $c$-axis, and perpendicular to the $c$-axis. The largest residual point symmetry consistent with this observation is 222, containing three perpendicular two-fold axes. For instance, if local structural distortions lower the crystal symmetry to the 222 point group, one of the three M-points is distinguished from the other two (which remain symmetry-related). This provides one possible explanation for the observed satellite structure, and is additionally consistent with a small-$q$ expansion akin to that of Section~\ref{section:small-q}, taking into account the lowered structural symmetry. Fig.~\ref{fig:domains_symmetry}(c,d) show the measured and inferred peak positions in this lower symmetry case. The dashed lines correspond to in-plane two-fold axes in the 222 group.



The satellites in Fig.~\ref{fig:domains_symmetry}(c) may be described by noticing that 
\begin{equation}
    \frac{\bm{q}_{\text{M}'}}{|\bm{q}_{\text{M}'}|} = R_c\left(\frac{2\pi}{6} + \alpha\right) \frac{\bm{q}_{\text{M}}}{|\bm{q}_{\text{M}}|},
\end{equation}
the direction of the satellite at M$'$ forms a relative angle $\alpha$ with the transverse direction at M$'$. Here $R_{\bm{c}}(\phi)$ denotes a rotation by $\phi$ about the $c$-axis. The satellites at M$''$ are obtained from those at M$'$ by the residual in-plane two-fold rotations. The satellites in Fig.~\ref{fig:domains_symmetry}(d) may be described by 
\begin{equation}
    \frac{\bm{q}_{\text{M}'}}{|\bm{q}_{\text{M}'}|} = R_c\left(\frac{2\pi}{6} - \alpha\right) \frac{\bm{q}_{\text{M}}}{|\bm{q}_{\text{M}}|}.
\end{equation}
Notably, in this case, the relative angle with the transverse direction at M$'$ is $-\alpha$. The magnitude of this relative angle $\alpha$ is observed to be the same across samples S1-S3, with a value of $\alpha \approx \frac{\pi}{6}$.

Because \cns is a chiral crystal, we propose that it is possible that there are two chiral structural domains related by inversion or reflections (which are \emph{not} symmetries of the \cns structure). In passing to a lower symmetry group due to structural distortions, the chirality of a domain is unchanged. The difference in the sign of this relative angle $\pm\alpha$ may be attributed to chiral domains of the crystalline structure, as mirror transformations relate the rotations $R_{c}\left(\frac{2\pi}{6} \pm \alpha\right)$.

The lower structural symmetry of \cns does not restrict the possible values for the angle $\alpha$. Although measurements indicate the slanted satellites point along the high-symmetry directions of the undistorted lattice, this direction is not enforced by the distorted crystal structure. If this angle deviates even slightly from $\frac{\pi}{6},$ the time-reversal translation symmetry prohibiting the AHE will be broken. Even if the slanted domains do not allow an AHE, our symmetry analysis suggests that the transverse domains are present in all samples, thus enabling a finite AHE. Future measurements on ultra-thin samples enabling REXS in a transmission geometry to access M, M$^{\prime}$, and M$^{\prime \prime}$ reflections in a single measurement or ultra-high resolution neutron diffraction measurements that can measure over many more Brillouin zones are required to experimentally confirm these proposed domain structures. The reduced structural symmetry and corresponding domain structure may also show strong signatures in optical dichroism and anisotropic transport measurements if single domains can be isolated.

We emphasize that the preceding analysis is a possible explanation for the experimental results under the assumption that there are no fundamental differences between samples S1-S3. This assumption is consistent with their similar heat capacity and magnetization measurements \cite{Note1}, and we find that this is possible if there exists a structural symmetry breaking.

\section{Discussion} \label{section:discussion}

We find that \cns exhibits a unique $2Q$ magnetic structure with staggered scalar spin chirality that is distinct from the reported tetrahedral $3Q$ structure \cite{takagi2023spontaneous,dong2024simple}. While our findings are fully consistent with the polarized neutron diffraction data \cite{takagi2023spontaneous}, a more than order-of-magnitude improved momentum space resolution reveals long-wavelength helical modulations that were not accessible to the neutron experiments and an analysis of the circular and linear polarization dependent magnetic REXS intensity rules out the tetrahedral $3Q$ order and uniform scalar spin chirality in \cns{}. Furthermore, the efficiency of REXS enabled measurements across many samples to reveal a subtle sample and domain dependence of the helical magnetic wavevector.

Although we identified a lowering of the structural symmetry from the observed magnetic domains, we cannot determine its origin. Given that the magnetic reflections observed in a given sample had intensities and orientations that were constant across many heating and cooling cycles, the domain structures are likely pinned by residual strain from the crystal growth, structural domain walls, or defect correlations that are quenched into the sample at high temperatures \cite{lim2024magnetochiral,du2021topological,horibe2014color,goodge2023consequences,choi2009giant}.

We found that a model Hamiltonian incorporating four-spin magnetic interactions can naturally account for the observed $2Q$ chiral stripe ordering in \cns. These four-spin interactions are essential to stabilize the long-wavelength helical modulation that break symmetries otherwise precluding the AHE in \cns. While four-spin interactions play a crucial role in the generation of AHE in both \cts~\cite{park2023tetrahedral,takagi2023spontaneous} and \cns, the mechanisms are completely different. In the former, the four-spin interaction induces a non-coplanar triple-$\bm{Q}$ tetrahedral ordering that produces a uniform scalar spin chirality. In the latter, the four-spin interaction induces an long-wavelength helical modulation of the commensurate order. The resulting non-uniform scalar chirality does not directly produce an AHE via the real-space Berry curvature, but breaks sufficient symmetries to allow an AHE. It is notable then, that despite the different underlying mechanisms, the resulting Hall response is extremely large and of comparable magnitude in both compounds \cite{ghimire2018large, park2022field}. Such a common phenomenology suggests that four-spin interactions, which can be significant in metallic magnets~\cite{batista2016frustration}, can play a fundamental role in the emergence of AHE via the generation of complex magnetic textures that significantly reduce the magnetic symmetry group relative to the paramagnetic crystal symmetry group.

Our work further shows how four-spin interactions in metallic magnets can act generally to engender more complex and potentially tunable electronic response, as a non-uniform scalar spin chirality is expected to influence electronic transport. For instance a finite local chirality can generate nonlinear or nonreciprocal transport \cite{hayami2022nonreciprocal,hayami2022nonlinear} which has been recently observed in \cns \cite{mi2023order}. The tunability of the magnetic structure demonstrated by pressure dependence \cite{popcevic2023electronic} and sample dependent magnetism in \cns \cite{mangelsen2021interplay,zhang2023chiral,tanaka2022large} and \cts \cite{park2024composition} together with the similarities of electronic structure between these two materials with different magnetic structures hints at the exciting possibility of controlling the magnetism in the intercalated TMDs to tune between different multi-$\bm{Q}$ orderings with a suitably chosen perturbation.

In summary, we have discovered a $2Q$ non-coplanar magnetic structure in \cns exhibiting a staggered scalar spin chirality $\chi_s$. Such a magnetic structure can be naturally explained by four-spin or biquadratic exchange in addition to isotropic Heisenberg exchange. The complex domain structure opens up possibilities for realizing and controlling nontrivial transport phenomena in metallic antiferromagnets. Future work is needed to fully understand the asymmetric domain structure and its role in the transport properties.

\section*{Acknowledgements}
Work at Brown University was supported by the U.S. Department of Energy, Office of Science, Office of Basic Energy Sciences, under Award Number DE-SC0021265. This work was carried out with the support of Diamond Light Source, beamline I10 under proposal numbers MM30765 and MM30768. We thank Mark Sussmuth for technical support at I10. CDB and S-S.Z. were supported by the  U.S.~Department of Energy, Office of Science, Office of Basic Energy Sciences, under Award Number DE-SC0022311. This work was  performed in part at the Aspen Center for Physics, which is supported by National Science Foundation grant PHY-2210452. Work at the University of California, Irvine was supported by the NSF through grant DMR-2142554.



\bibliography{bib}

\end{document}


\title{\texorpdfstring{Supplementary Material for Double-$Q$ chiral stripe order in the anomalous Hall antiferromagnet \cns}{}}

\author{Ben Zager}
\affiliation{Department of Physics, Brown University, Providence, Rhode Island 02912, USA}

\author{Shang-Shun Zhang}
\affiliation{ Department of Physics and Astronomy, University of Tennessee, Knoxville, TN 37996, USA}

\author{Hana Schiff}
\affiliation{Department of Physics and Astronomy, University of California, Irvine, California
92697, USA}

\author{Raymond Fan}
\affiliation{Diamond Light Source Ltd., Harwell Science and Innovation Campus, Didcot OX11 0DE, United Kingdom}

\author{Paul Steadman}
\affiliation{Diamond Light Source Ltd., Harwell Science and Innovation Campus, Didcot OX11 0DE, United Kingdom}

\author{Cristian D. Batista}
\affiliation{ Department of Physics and Astronomy, University of Tennessee, Knoxville, TN 37996, USA}
\affiliation{Neutron Scattering Division, Oak Ridge National Laboratory, Oak Ridge, TN 37831, USA }

\author{Kemp W. Plumb}
\email[Corresponding author:]{kemp_plumb@brown.edu}	
\affiliation{Department of Physics, Brown University, Providence, Rhode Island 02912, USA}

\date{\today}

\maketitle


\vspace{0mm}

\begin{figure}[h!]
    \includegraphics[scale=0.9]{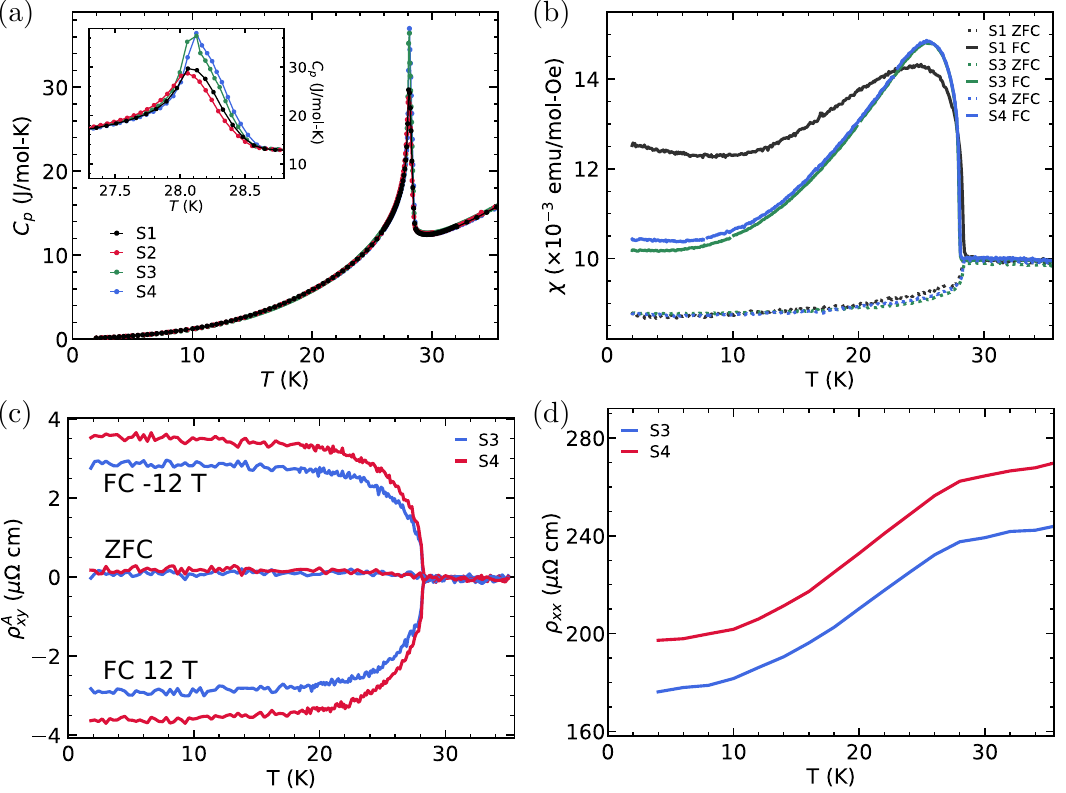}
    \caption{
    (a) Heat capacity vs. temperature for samples 1-4.
    (b) Magnetic susceptibility vs. temperature measured with a 0.1~T field with both zero-field-cooling and field-cooling for samples 1,3, and 4.
    (c) Anomalous Hall resistivity $\rho_{xy}$ vs. $T$ for samples 3 and 4.
    (d) Longitudinal resistivity $\rho_{xx}$ vs. $T$ for samples 3 and 4.
    }
    \label{fig:Cp_M}
\end{figure}

\section{Sample characterization}

Heat capacity, magnetic susceptibility, and electrical transport measurements were carried out using a Quantum Design Physical Property Measurement System (PPMS). Susceptibility was measured using the vibrating sample mount (VSM) with the magnetic field along the sample $c$-axis. Electrical transport was measured with the van der Paux method using a 2~mA excitation current. 50~$\mu$m Au wires were attached to the samples using silver paint.

Fig. \ref{fig:Cp_M}(a) shows the heat capacity of S1-S4. All samples display a clear second-order transition at $T_N = 28.6$~K. The entropy loss at the transition is slightly greater for S3 and S4, indicating possible intrinsic differences between samples.
Fig. \ref{fig:Cp_M}(b) shows the magnetic susceptibility vs. temperature for S1, S3, and S4. S2 was too small to produce a sufficient signal. A constant background $\chi_0$ was subtracted from each sample, with $\chi_0 = -2$, 13, and 2.3 $\times 10^{-3}$ emu/mol-Oe for S1, S3, and S4 respectively. All samples show a transition at $T_N = 28.6$~K with a large splitting between field (FC) and zero field cooling (ZFC), consistent with previous reports \cite{ghimire2018large,tanaka2022large,lu2022understanding,zhang2023chiral,gu2025probing}. Fig. \ref{fig:Cp_M}(c) shows the anomalous Hall conductivity $\sigma_{xy}^A = \rho_{xy}^A/[\rho_{xx}^2+(\rho_{xy}^A)^2]$ vs. temperature for S3 and S4, measured for zero-field cooling and $\pm$12~T field-cooling. Fig. \ref{fig:Cp_M}(d) shows the longitudinal resistivity for S3 and S4.

\section{Sample preparation for resonant elastic x-ray scattering measurements}

\begin{figure}[h!]
   \includegraphics[scale=0.85]{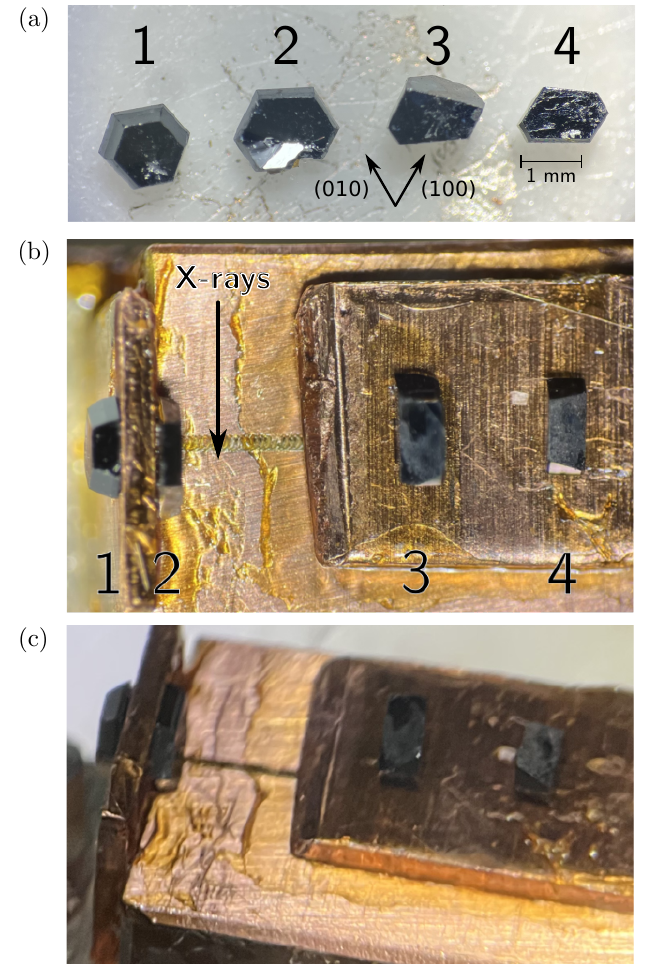}
   \caption{
   (a) Microscope image of samples 1-4 (S1-S4).
   (b) Image of samples 1-4 mounted for the RASOR endstation Diamond I10, showing the beam direction at $\theta = 0\degree$.
   (c) Same as (b) from a different angle.
   }
   \label{fig:samples}
\end{figure}

Images of the single crystals used for our measurement are shown in Fig.~\ref{fig:samples}. The samples are naturally mirror faceted; however, we note the roughness of the scattering surface of S4 seen in \ref{fig:samples}(b) and (c), which will be discussed below.
Samples were pre-aligned on copper bars and subsequently mounted on the copper sample holder using GE varnish. The temperature sensor was located on the coldfinger, at the base of the sample holder closest to S4. To fit the four samples on the sample holder, they were mounted in different geometries, with differing areas in contact with the holder. Thus, each sample experiences a different cooling efficiency.

\section{Reciprocal space maps for samples 1-4}

\begin{figure}[h!]
   \includegraphics{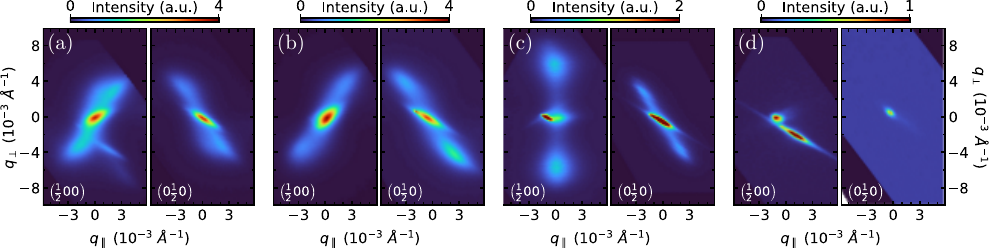}
   \caption{Reciprocal space maps of the magnetic scattering at $(\frac{1}{2}00)$ and $(0\frac{1}{2}0)$ in the $(HK0)$ plane integrated over 0.003 r.l.u. along the $L$ direction for (a) S1, (b) S2, (c) S3, and (d) S4.}
   \label{fig:qmap_all}
\end{figure}

We measured reciprocal space maps of the magnetic scattering around the $(\tfrac{1}{2}00)$ and $(0\tfrac{1}{2}0)$ wavevectors in four samples (S1-S4), as shown in Fig. \ref{fig:qmap_all}. Panels~(a) and (b) show the maps for S1 and S2 respectively, which show the same structure. As discussed in the main text, the modulations in the two $\bm{Q}_0$-domains are symmetry-related, but each is missing the other symmetry-allowed $\bm{q}$ modulation. S1 $(\tfrac{1}{2}00)$ shows some weak overlapping intensity below the main peak that may come from a different grain with satellites along the other symmetry-allowed direction. Panel~(c) shows the map for S3, as in Fig.~1(a) of the main text. In this sample, the modulations in the two $\bm{Q}_0$-domains are not symmetry-related. Panel~(d) shows the map for S4. This sample shows $\bm{Q}_0$ magnetic reflections, with two distinct grains appearing at $(\tfrac{1}{2}00)$, but no satellite reflections. We attribute this to the surface roughness, which is evident in Fig.~\ref{fig:samples}. This roughness likely disrupts the incommensurate order via internal strain or reduced structural correlation lengths, consistent with previous reports on itinerant magnets \cite{fawcett1988sdw,boekelheide2007sdw}. This may also explain the weaker $\bm{Q}_0$ signal (note the different scale bars) and increased diffuse background at $(0\tfrac{1}{2}0)$, which is the more grazing angle. Although the characterization presented below indicates structural correlation lengths in S4 comparable to the other samples, those measurements use the 2nd harmonic of the beam, which has more than four times greater penetration depth than the fundamental.


\section{Correlation lengths}

\begin{figure}[h!]
    \includegraphics[scale=0.7]{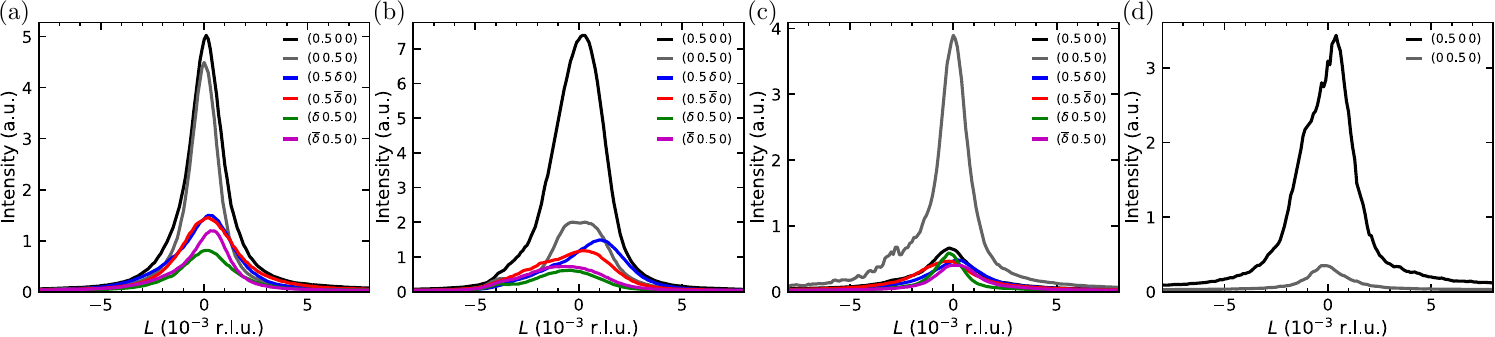}
    \caption{
    Line cuts through the magnetic peaks along the $L$ direction integrated over 0.005~r.l.u. along the $H$ and $K$ directions in (a) S1, (b) S2, (c) S3, and (d) S4.
    }
    \label{fig:xi_c}
    \end{figure}

\begin{figure}[h!]
    \includegraphics[scale=0.7]{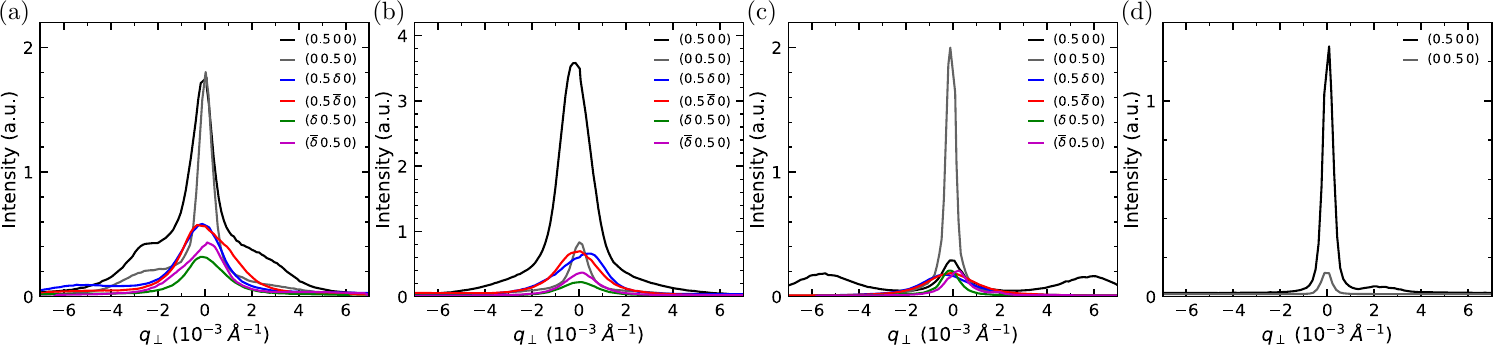}
    \caption{Line cuts through the magnetic peaks along the in-plane transverse direction, integrated over 0.003~r.l.u. along the $L$ direction and 0.005~r.l.u. along the in-plane longitudinal direction, in (a) S1, (b) S2, (c) S3, and (d) S4.
    }
    \label{fig:xi_ab}
\end{figure}

\begin{figure}[h!]
    \includegraphics[scale=0.7]{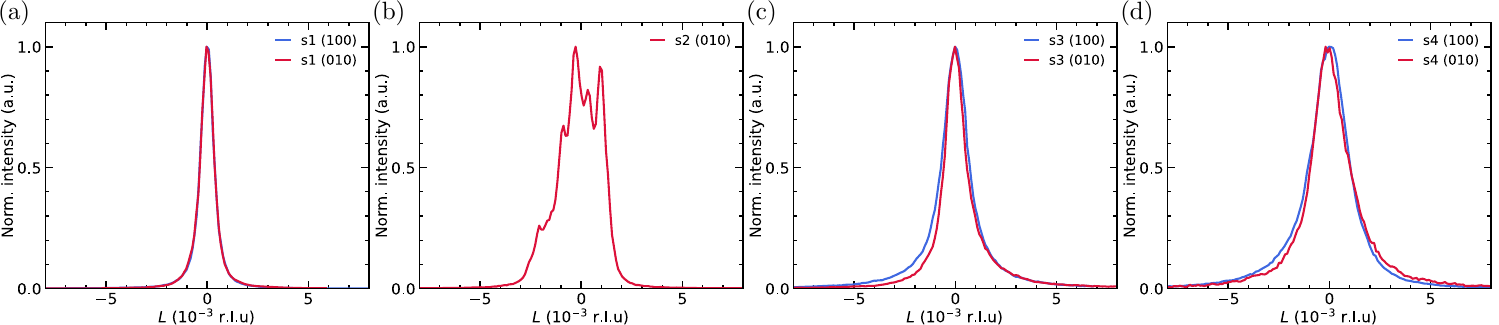}
    \caption{Line cuts through the structural peaks along the $L$ direction, integrated over 0.005~r.l.u along the $H$ and $K$ directions, with maximum intensity normalized to 1 in (a) S1, (b) S2, (c) S3, and (d) S4, measured on the second harmonic (1557 eV) at 30~K.
    }
    \label{fig:xi_c_30K}
\end{figure}

\begin{figure}[h!]
    \includegraphics[scale=0.7]{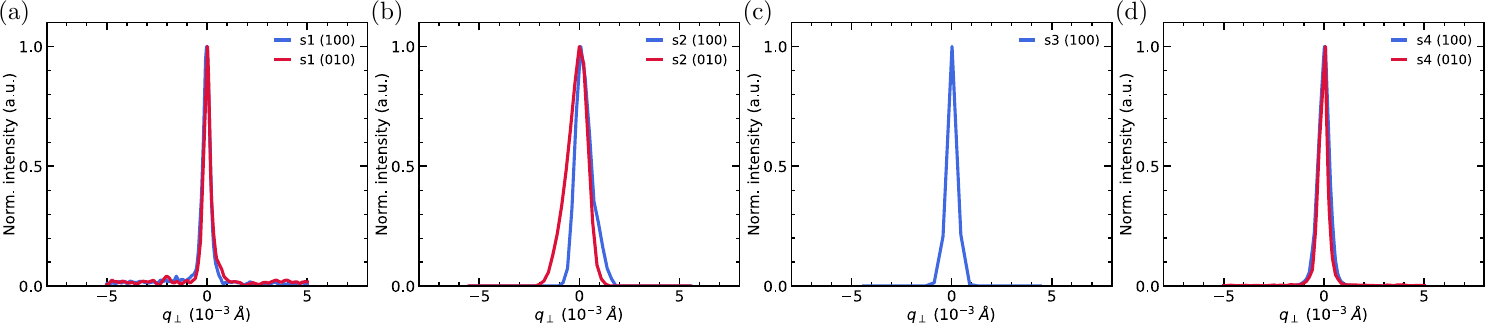}
    \caption{Line cuts through the structural peaks along the in-plane transverse direction, integrated over 0.003~r.l.u. along the $L$ direction and 0.005~r.l.u. along the in-plane longitudinal direction, in (a) S1, (b) S2, (c) S3, and (d) S4, measured on the second harmonic (1557~eV) at 30~K.
    }
    \label{fig:xi_ab_30K}
\end{figure}

Fig.~\ref{fig:xi_c} shows line cuts along the $L$ direction for each main and satellite peak in all samples. We fit a Lorentzian function to obtain an out-of-plane correlation length $\xi_c$. Fig. \ref{fig:xi_ab} shows in-plane transverse line cuts across each peak, which we fit to obtain an in-plane correlation length $\xi_{ab}$. For the satellites, the transverse direction is taken with respect to $\bm{q}$, rather than $\bm{Q}_0\pm\bm{q}$. Fig.~\ref{fig:xi_c_30K} and Fig.~\ref{fig:xi_ab_30K} show cuts through the structural peaks measured on the second harmonic (1557~eV) at 30~K along the $L$ direction and in-plane transverse directions respectively. The structural and magnetic correlation lengths extracted from these fits are summarized in tables \ref{tab:S124_corrLength} and \ref{tab:S3_corrLength} below. These (100)-type structural Bragg peaks arise from the superlattice of intercalated Co ions within the NbS$_2$ host layers and the large correlation lengths indicate a well-ordered triangular lattice of Co ions \cite{ueno2005xray} in all samples studied. 

The magnetic correlation lengths for the incommensurate structure are on the order of 100 nm, taken from transverse cuts along the satellite peaks. These peaks are nearly isotropic, showing similar width in any direction. This confirms that the magnetic structure we observed is intrinsic to the bulk. If the incommensurate structure was confined to the surface, these peaks would show a truncation rod-like elongation along the direction of the surface normal \cite{watson2000resonant,ran2021creation}. In addition, we find that $\bm{q}$ is always along a high-symmetry direction, independent of the surface orientation. 

\renewcommand{\arraystretch}{1.5}
\begin{table}[h!]
  \centering
  \caption{Measured correlation lengths for S1, S2, and S4.}
  \begin{tabular}{c c c c c c c c c}
  \hline \hline
  Peak  & $(100)$ & $(010)$ & $(\tfrac{1}{2}00)$ & $(\tfrac{1}{2}\delta0)$ & $(\tfrac{1}{2}\overline{\delta}0)$ & $(0\tfrac{1}{2}0)$ & $(\delta\tfrac{1}{2}0)$ & $(\overline{\delta}\tfrac{1}{2}0)$ \\
  \hline
  {\bf Sample 1} &&&&&&&&\\
  \hline
  $\xi{ab}$ (nm) & 563(8) & 570(9) & 151(14) & 93(3) & 85(2)   & 247(7) & 107(1) & 102(2) \\ 
  $\xi_{c}$ (nm)  & 571(5) & 563(5) & 212(1)  & 135(1) & 133(1) & 265(2) & 143(1) & 172(1) \\ 
  \hline
  {\bf Sample 2} &&&&&&&&\\
  \hline 
  $\xi_{ab}$ (nm) & 274(14) & 216(12) & 136(2) & 197(5) & 90(2)  & 99(2)  & 107(2) & 122(2) \\ 
  $\xi_{c}$ (nm)  & -       & 182(6)  & 159(2) & 139(2) & 112(2) & 98(2)  & 110(2) & 97(1) \\
  \hline
  {\bf Sample 4} &&&&&&&&\\
  \hline
  $\xi_{ab}$ (nm) & 394(8) & 474(11) & 469(40) & - & - & 352(37) & - & - \\ 
  $\xi_{c}$ (nm)  & 195(2) & 199(2)  & 146(3)  & - & - & 151(4)  & - & - \\ 
  \hline\hline
  \end{tabular}
  \label{tab:S124_corrLength}
\end{table}
\begin{table}[h!]
  \centering
  \caption{Measured correlation lengths for S3.}
  \begin{tabular}{c c c c c c c c c}
  \hline\hline
  Peak  & $(100)$ & $(010)$ & $(\tfrac{1}{2}00)$ & $(\tfrac{1}{2}\!+\!\delta,-2\delta,0)$ & $(\tfrac{1}{2}\!-\!\delta,2\delta,0)$ & $(0\tfrac{1}{2}0)$ & $(\delta\tfrac{1}{2}0)$ & $(\overline{\delta}\tfrac{1}{2}0)$ \\
  \hline
  $\xi_{ab}$ (nm) & 274(14) & 216(12) & 136(2) & 197(5) & 90(2)  & 99(2) & 107(2) & 122(2) \\ 
  $\xi_{c}$ (nm)  & 268(1)  & 321(2)  & 120(1) & 111(1) & 112(2) & 224(3) & 231(3) & 153(2) \\
  \hline\hline
  \end{tabular}
  \label{tab:S3_corrLength}
\end{table}



\section{Temperature dependence of magnetic scattering}
\begin{figure}[h!]
    \includegraphics[scale=0.8]{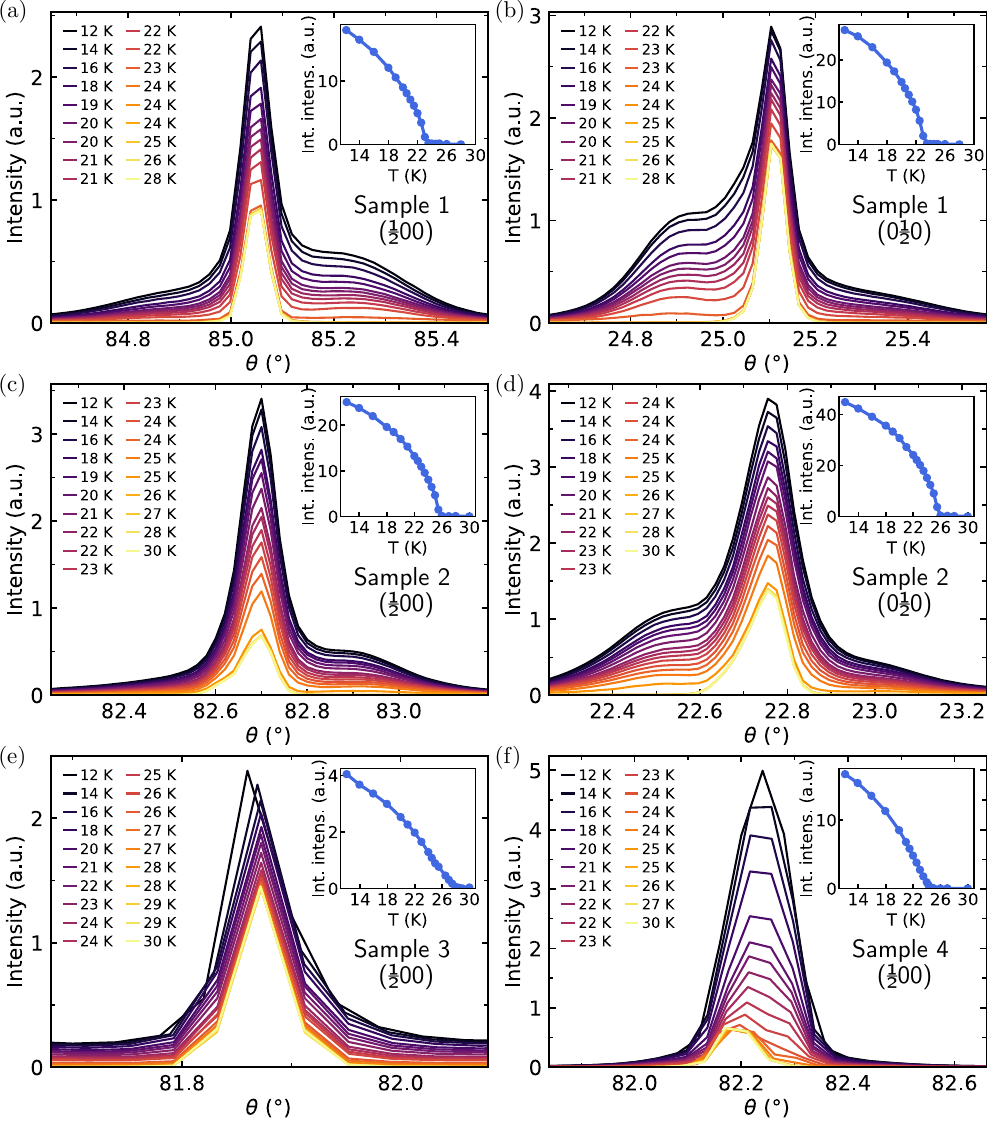}
    \caption{Temperature dependence of magnetic scattering measured from rocking scans around $\bm{Q}_0$ in $\pi$ polarization on the point detector for 
    (a) S1 $(\frac{1}{2}00)$,
    (b) S1 $(0\frac{1}{2}0)$,
    (c) S2 $(\frac{1}{2}00)$,
    (d) S2 $(0\frac{1}{2}0)$,
    (e) S3 $(\frac{1}{2}00)$,
    (f) S4 $(\frac{1}{2}00)$.
    }
    \label{fig:Tdep}
\end{figure}

Fig. \ref{fig:Tdep} shows the temperature dependence of the magnetic scattering for samples 1-4 measured from rocking scans on the point detector in $\pi$ polarization. Above the transition, only the $(100)$ or $(010)$ structural peaks from the second harmonic of the beam remain. The insets show the integrated intensity vs. temperature with the structural peak intensity subtracted. Although $T_N$ as measured through susceptibility and heat capacity is consistent across all samples, the apparent $T_N$ in the REXS data varies between samples. This variation arises from a combination of beam heating and variable cooling efficiency for samples mounted on different locations on the sample holder as shown in  Fig.~\ref{fig:samples}.


\section{Description of the Magnetic structure}

Our data reveal that the magnetic structure of \cns is comprised of two propagation vectors $\bm{Q}_0$ and $\bm{q}$. There are two possible magnetic structures that may be consistent with this observation. In the main text, we consider the case with magnetic propagation vectors $\pm\bm{Q}_0$ and $\pm(\bm{Q}_0\!\pm\!\bm{q})$. The second possible scenario is a structure with propagation vectors $\bm{Q}_0$ and $\bm{q}$, giving rise to higher harmonic peaks at $\bm{Q}_0\!\pm\!\bm{q}$. We cannot directly rule out low-angle peaks at $\bm{q}$ that were inaccessible in our scattering geometry. However, we did not observe any $\bm{Q}_0\!\pm\! n\bm{q}$ for $n\!>\!1$ higher harmonics, which would be expected in this scenario. Furthermore, neutron experiments have not reported any evidence for $\bm{q} \!\approx\! 0$ magnetic scattering \cite{parkin1983magnetic, lu2022understanding}. We thus rule out the second scenario and consider the structure with magnetic propagation vectors $\pm\bm{Q}_0$ and $\pm(\bm{Q}_0\!\pm\!\bm{q})$. The real-space structure is given by 
\begin{equation} \label{eq:magnetic_structure}
\begin{aligned}
    \bm{S}_n^{}(\bm{r}_j)
    &= S\cos\phi_n \cos[(\bm{Q}_0+\bm{q}) \cdot \bm{r}_j+\psi_n] \hat{\bm{u}}_n \nonumber \\
    &+ S\sin\phi_n \cos(\bm{Q}_0 \cdot \bm{r}_j) \hat{\bm{v}}_n \nonumber \\
    &+ \chi S \cos\phi_n \sin[(\bm{Q}_0+\bm{q}) \cdot \bm{r}_j+\psi_n] \hat{\bm{w}}_n,
\end{aligned}
\end{equation}
where $n\!=\!1,2$ labels the sublattices at 2 Co sites in the unit cell, $\chi\! =\! \pm 1$ is the helix chirality, $\phi_n$ and $\psi_n$ are the phases on sublattice $n$ for $\bm{Q}_0$ and $\bm{Q}_0\! \pm\! \bm{q}$ respectively, and $\hat{\bm{u}}_n$, $\hat{\bm{v}}_n$, and $\hat{\bm{w}}_n$ are unit vectors, assumed to be orthogonal to maintain a constant moment size. Eq.~\ref{eq:magnetic_structure} can also be expressed as a generalized form of the $2Q$ structure considered in \cite{heinonen2022magnetic}
\begin{equation}
\begin{aligned}
    \bm{S}_n^{}(\bm{r}_j)
    &= \cos(\bm{q}\cdot\bm{r}_j + \psi_n)\cos(\bm{Q}_0\cdot\bm{r}_j + \phi_n)\hat{\bm{u}}_n \\
    &+ \sin(\bm{Q}_0\cdot\bm{r}_j + \phi_n)\hat{\bm{v}}_n \\
    &+ \cos(\bm{q}\cdot\bm{r}_j + \psi_n + \tfrac{\pi}{2}\chi)\cos(\bm{Q}_0\cdot\bm{r}_j + \phi_n) \hat{\bm{w}}_n.
\end{aligned}
\end{equation}

The magnetic moment on site $j$ can be expressed as 
\begin{equation}
    \bm{S}(\bm{r}_j) = \sum\limits_{\bm{Q}}\bm{S}_{\bm{Q}}^{}e^{i\bm{Q}\cdot\bm{r}_j}
\end{equation}
where $\bm{S}_{\bm{Q}}^{}$ are the Fourier components of the magnetic moments, which, for this structure are given by

\begin{eqnarray}
    \bm{S}_{\bm{Q}_0,n}^{} &=& \sqrt{N}S\sin\phi_n \hat{\bm{v}}_n, \nonumber \\
    \bm{S}_{\bm{Q}_0\pm\bm{q},n}^{} &=& \sqrt{N}{S\over 2} \cos\phi_n e^{i\psi_n} (\hat{\bm{u}}_n \pm i\chi\hat{\bm{w}}_n),
\end{eqnarray}
where $\bm{S}_{\bm{Q}_0,n}^{} = \bm{S}_{-\bm{Q}_0,n}^* = \bm{S}_{\bm{Q}_0,n}^*$, $\bm{S}_{\bm{Q}_0\pm\bm{q},n}^{} = \bm{S}_{-\bm{Q}_0\mp\bm{q},n}^*$, and $N$ is the number of unit cells.


The precise value of $\phi_n$ affects the real-space spin structure, ruling out $\phi_n\!=\!\frac{\pi}{2}n$ ($n=0,1,2,\ldots$) for which either the commensurate or incommensurate components of the magnetic structure vanish. The precise value of $\psi_n$ does not affect the real-space structure, since $\bm{q}$ is incommensurate with the lattice. However, the relative phase $\Delta\psi = \psi_2 - \psi_1$ does affect the real-space structure. The unit vectors defining the moment orientations are
\begin{equation}
\begin{aligned}
    \hat{\bm{u}}_n &= \hat{\bm{x}}\cos\mu + \hat{\bm{y}}\sin\mu \\
    \hat{\bm{v}}_n &= -\hat{\bm{x}}\sin\mu\cos\nu_n + \hat{\bm{y}}\cos\mu\cos\nu_n + \hat{\bm{z}}\sin\nu_n \\
    \hat{\bm{w}}_n &= \hat{\bm{x}}\sin\mu\sin\nu_n - \hat{\bm{y}}\sin\nu_n\cos\mu + \hat{\bm{z}}\cos\nu_n
\end{aligned}
\end{equation}
for $n=1,2$, where $\hat{\bm{x}}, \hat{\bm{y}}, \hat{\bm{z}}$ are Cartesian coordinates with $\hat{\bm{x}}$ along $a$ and $\hat{\bm{y}}$ in the $a$-$b$ plane of the crystallographic cell. $\mu$ defines the in-plane angle relative to the $x$-axis, $\nu_n$ defines the out-of-plane canting angle where we assume $\nu = \nu_1 = -\nu_2$.

\section{Calculation of the resonant magnetic x-ray scattering intensity}

\cns crystallizes in the chiral $P6_322$ space group (182) with lattice parameters $a=5.75$~\AA{} and $c=11.89$~\AA{}. In the fully ordered structure, the Co$^{2+}$ ions exclusively occupy either the $2b$, $2c$, or $2d$ site \cite{mangelsen2021interplay}. Here we consider consider the $2c$ sites $(\tfrac{1}{3},\tfrac{2}{3},\tfrac{1}{4})$ and $(\tfrac{2}{3},\tfrac{1}{3},\tfrac{3}{4})$ to be occupied, but this choice does not affect the results. We use the lattice vectors $\bm{a}_1 = (a,0,0)$, $\bm{a}_2 = \tfrac{a}{2}(-1,\sqrt{3},0)$, and $\bm{a}_3 = (0,0,c)$, which give reciprocal lattice vectors $\bm{b}_1 = \tfrac{2\pi}{a}(1,\tfrac{1}{\sqrt{3}},0)$, $\bm{b}_2 = \tfrac{2\pi}{a}(0,\tfrac{2}{\sqrt{3}},0)$, and $\bm{b}_3 = \tfrac{2\pi}{c}(0,0,1)$. The REXS intensity is $I_{\bm{\epsilon}'\bm{\epsilon}}(\bm{Q}) \propto |F_{\bm{\epsilon}'\bm{\epsilon}}(\bm{Q})|^2$, where
\begin{equation}
    F_{\bm{\epsilon}'\bm{\epsilon}}(\bm{Q}) = \sum\limits_je^{i\bm{Q}\,\cdot\,\bm{r}_j}f_j(E)
\end{equation}
is the structure factor for incident (outgoing) polarization vector $\bm{\epsilon}$ ($\bm{\epsilon}'$), which sums over all Co sites $j$, and $E$ is the x-ray energy. In the dipole approximation, the resonant form factor at atomic site $j$ is
\begin{equation}
    f_j = (\bm{\epsilon}'^*\cdot\bm{\epsilon})F^{(0)}
    + (\bm{\epsilon}'^*\times\bm{\epsilon})\cdot\bm{S}_jF^{(1)}
    + (\bm{\epsilon}'^*\cdot\bm{S}_j)(\bm{\epsilon}\cdot\bm{S}_j)F^{(2)}
\end{equation}
where $\bm{S}_j$ is the direction of the magnetic moment on the site and $F^{(i)}$ are the energy-dependent resonant scattering factors \cite{hill1996resonant}. We observed no intensity in the $\sigma$-$\sigma$ channel, so we let $F^{(0)}=0$. The third term gives rise to second-order satellite reflections, which we did not observe, so we let $F^{(2)}=0$. We thus have
\begin{equation} \label{eq:sf}
    F_{\bm{\epsilon}'\bm{\epsilon}}(\bm{Q}) \propto 
    (\bm{\epsilon}'^*\times\bm{\epsilon})\cdot\bm{M}(\bm{Q})
\end{equation}
where $\bm{M}(\bm{Q}) = \bm{S}_{\bm{Q},1}e^{i\bm{Q}\,\cdot\,\bm{r}_1} + \bm{S}_{\bm{Q},2}e^{i\bm{Q}\,\cdot\,\bm{r}_2}$ is the magnetic structure factor. For the main magnetic reflection at $\bm{Q}_0 = (\tfrac{1}{2}00)$,
\begin{equation}
    \bm{M}(\bm{Q}_0) = e^{i\frac{\pi}{3}}\sin{\phi_1}\hat{\bm{v}}_1 
    + e^{i\frac{2\pi}{3}}\sin{\phi_2}\hat{\bm{v}}_2,
\end{equation}
where we ignore the $\sqrt{N}S$ factor. For the $\bm{Q}_0\pm\bm{q} = (\tfrac{1}{2}00) \pm (0\delta 0)$ satellites,
\begin{equation}
    \bm{M}(\bm{Q}_0\pm\bm{q}) = \frac{1}{2}\bigg[
    \cos\phi_1e^{i(\frac{\pi}{3} \pm \frac{4\pi}{3}\delta + \psi_1)}(\hat{\bm{u}}_1 \pm i\chi\hat{\bm{w}}_1)
    + \cos\phi_2e^{i(\frac{2\pi}{3} \pm \frac{2\pi}{3}\delta + \psi_2)}(\hat{\bm{u}}_2 \pm i\chi\hat{\bm{w}}_2)
    \bigg].
\end{equation}
For the $\bm{Q}_0\pm\bm{q} = (\frac{1}{2}00) \pm (\delta,-2\delta,0)$ satellites,
\begin{equation}
    \bm{M}(\bm{Q}_0\pm\bm{q}) = \frac{1}{2}\Bigg[
    \cos\phi_1e^{i(\frac{\pi}{3} \pm 2\pi\delta + \psi_1)}(\hat{\bm{u}}_1 \pm i\chi\hat{\bm{w}}_1)
     + \cos\phi_2e^{i(\frac{2\pi}{3} + \psi_2)}(\hat{\bm{u}}_2 \pm i\chi\hat{\bm{w}}_2)
    \bigg].
\end{equation}
Similar expressions can be obtained for $\bm{Q}_0 = (0\frac{1}{2}0)$ and the corresponding satellites.

\begin{equation}
    \bm{M}(\bm{Q}_0\pm\bm{q}) = \frac{1}{4}\Big[
    ie^{i(\pm\frac{4\pi}{3}\delta+\phi_1)}\hat{\bm{u}}_1 
\mp \chi e^{i(\pm\frac{4\pi}{3}\delta+\phi_1)}\hat{\bm{w}}_1 
  + ie^{i(\pm\frac{2\pi}{3}\delta+\phi_2)}\hat{\bm{u}}_2 
\mp \chi e^{i(\pm\frac{2\pi}{3}\delta+\phi_2)}\hat{\bm{w}}_2\Big]
\end{equation}
For the $\bm{q} = (\delta,-2\delta,0)$ satellites,
\begin{equation}
    \bm{M}(\bm{Q}_0\pm\bm{q}) = \frac{1}{4}\Big[
    ie^{i(\phi_1 \mp 2\pi\delta)}\hat{\bm{u}}_1
    \mp\chi e^{i(\phi_1 \mp 2\pi\delta)}\hat{\bm{w}}_1
    + ie^{i\phi_2}\hat{\bm{u}}_2
    \mp\chi e^{i\phi_2}\hat{\bm{w}}_2
    \Big]
\end{equation}

\subsection{Full linear polarization analysis (FLPA)}


Let $\hat{\bm{Q}}$ be the unit vector along the direction of the scattering vector $\bm{Q}$ and let $\hat{\bm{Q}}_{\perp}$ be a direction perpendicular to $\hat{\bm{Q}}$ within the $(HK0)$ scattering plane. The directions of the incident and scattered beam are given by
\begin{equation}
\begin{aligned}
    \hat{\bm{k}} &= \hat{\bm{Q}}_{\perp}\cos\theta + \hat{\bm{Q}}\sin\theta \\
    \hat{\bm{k}}' &= \hat{\bm{Q}}_{\perp}\cos\theta - \hat{\bm{Q}}\sin\theta.
\end{aligned}
\end{equation}
where $\theta$ is half of the Bragg angle \cite{haverkort2010symmetry}. We can represent the polarization in the basis $\bm{\sigma}$ (perpendicular to the scattering plane) and $\bm{\pi}$ (parallel to the scattering plane), given by
\begin{equation}
\begin{aligned}
    \bm{\sigma} &= \bm{\sigma}' = (\hat{\bm{k}}\times\hat{\bm{k}}')/\sin(2\theta) \\
    \bm{\pi} &= \hat{\bm{k}} \times \bm{\sigma} \\
    \bm{\pi}' &= \hat{\bm{k}}' \times \bm{\sigma}.
\end{aligned}
\end{equation}
The linear polarization vector is then given by
\begin{equation}
\begin{aligned}
    \bm{\epsilon} &= \bm{\sigma}\cos\alpha + \bm{\pi}\sin\alpha \\
    \bm{\epsilon}' &= \bm{\sigma}'\cos\eta + \bm{\pi}'\sin\eta
\end{aligned}
\end{equation}
The REXS intensity can then be calculated using Eq. \ref{eq:sf}. As expected for the $E1$-$E1$ contribution to magnetic scattering, the $\sigma$-$\sigma'$ intensity is always zero. The $\sigma$-$\pi'$ and $\pi$-$\sigma'$ intensity originates from the component of the moment in the crystallographic $a$-$b$ (basal) plane and the $\pi$-$\pi'$ intensity arises from the out-of-plane component. 

\begin{figure}[h!]
    \includegraphics[scale=1.0]{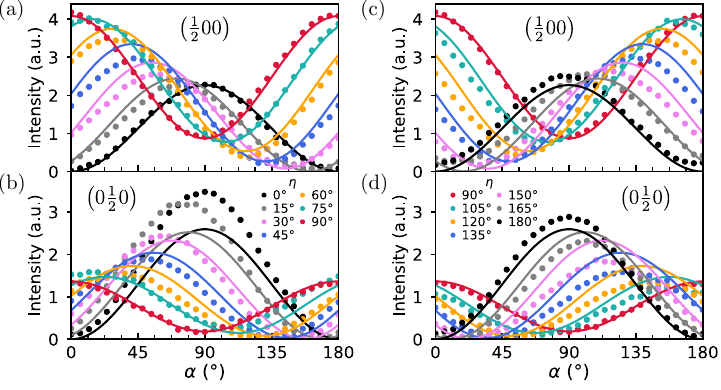}
    \caption{Individual FLPA scans and calculation from Fig. 4 of the main text. (a) and (b) show $\eta = 0\text{-}90^{\degree}$ for $(\tfrac{1}{2}00)$ and ($(0\tfrac{1}{2}0)$) respectively. (c) and (d) show $\eta = 90\text{-}180^{\degree}$ for $(\tfrac{1}{2}00)$ and $(0\tfrac{1}{2}0)$ respectively.
    }
    \label{fig:flpa_SI}
\end{figure}

\subsection{Circular dichroism}

For circularly polarized x-rays $\bm{\epsilon}_{\pm} = \tfrac{1}{\sqrt{2}}(\bm{\sigma} \mp i\bm{\pi})$, the intensity $I_{\pm} = I_{\pm\pm} + I_{\mp\pm}$ is \cite{mulders2010circularly}
\begin{equation}
    I_{\pm} = \frac{1}{2}(|F_{\sigma'\sigma}^{}|^2 + |F_{\sigma'\pi}^{}|^2 + |F_{\pi'\sigma}^{}|^2 + |F_{\pi'\pi}^{}|^2)
    \pm \text{Im}(F_{\sigma'\sigma}^{}F_{\sigma'\pi}^* + F_{\pi'\sigma}^{}F_{\pi'\pi}^*),
\end{equation}
The circular dichroism is 
\begin{equation}
    I_+ - I_- = 2\,\text{Im}(F_{\pi'\sigma}^{}F_{\pi'\pi}^*),
\end{equation}
where we have assumed $F_{\sigma'\sigma}^{} = 0$. Thus, a necessary condition for finite CD is a finite structure factor for both $\sigma$-$\pi'$ and $\pi$-$\pi'$. This requires both an in- and out-of-plane spin component, which indicates that the modulation is helix-like and confirms the FLPA results of a non-collinear commensurate component.

\begin{figure}[h!]
    \includegraphics{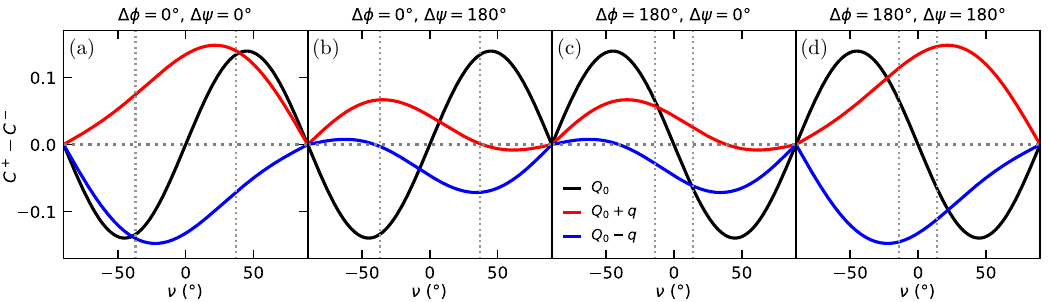}
    \caption{Dependence of the CD on the canting angle $\nu$ at $\bm{Q}_0 = (\tfrac{1}{2}00)$ and $\bm{Q}_0\pm(0\delta 0)$ for each of the four phase combinations with $\mu=109\degree$ and $\phi_1 = 45\degree{}$. The vertical dashed lines show the value of $\nu$ determined from FLPA for each choice of phases.}
    \label{fig:cd_calc}
\end{figure}

Fig.~\ref{fig:cd_calc} shows the calculated CD as a function of canting angle $\nu$ at $\mu = 109\degree$. The magnetic structure determined from FLPA is consistent with the CD-REXS results. The condition for CD at $\bm{Q}_0$ is finite structure factor for both $\sigma$-$\pi'$ and $\pi$-$\pi'$, which arises from the noncollinear moments with both in- and out-of-plane components. The CD at $\bm{Q}_0\pm\bm{q}$ requires a helical structure or a more complex chiral modulation. We note that there is a contribution to the CD from the structural peaks on second harmonic, which is forbidden for Thomson scattering. This is due to the imperfect circular polarization of the second harmonic, giving unequal combinations of $\sigma$ and $\pi$ polarization between $C+$ and $C-$ \cite{wang2012complete}.

\subsection{Calculations for the \texorpdfstring{triple-$Q$}{} order}

\begin{figure}[h!]
    \includegraphics[scale=0.85]{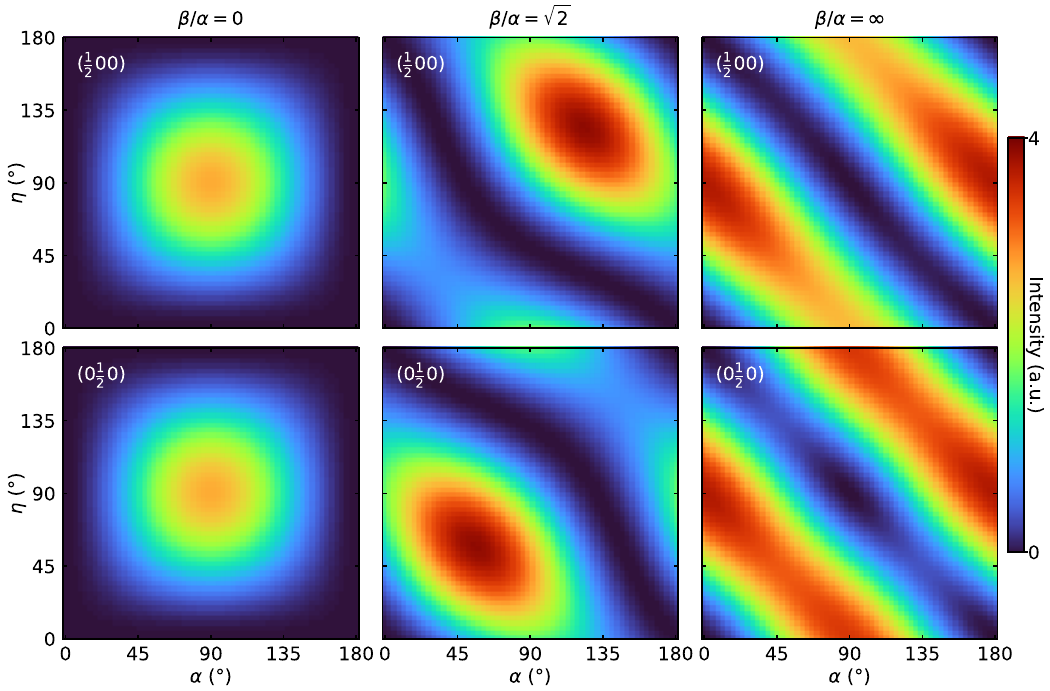}
    \caption{Calculated FLPA patterns for the tetrahedral $3Q$ state at ($\tfrac{1}{2}00$) (top) and ($0\tfrac{1}{2}0$) (bottom), for 
    $\beta/\alpha = 0$ (left), 
    $\beta/\alpha = \sqrt{2}$ (middle), and 
    $\beta/\alpha = \infty$ (right).
    } 
    \label{fig:flpa_3q}
\end{figure}

We have calculated the FLPA pattern for the tetrahedral $3Q$ structure for three different values of the parameter $\beta/\alpha$ \cite{park2023tetrahedral}, shown in Fig.~\ref{fig:flpa_3q}. As described in the main text, none of these cases can reproduce our measurements.

\section{Scalar spin chirality}

\begin{figure}[h!]
    \includegraphics[scale=1.2]{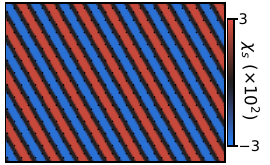}
    \caption{
    Scalar spin chirality $\chi_s^{\parallel}$ from the intra-sublattice plaquettes, summed over both sublattices for the structure $\bm{Q}_0\! =\! (\tfrac{1}{2}00)$ and $\bm{q}\! =\! (-\delta,2\delta,0)$, $\delta \!=\! 0.053$ r.l.u., $\phi \!=\! 45\degree$.
    }
    \label{fig:chi_IP}
\end{figure}

The scalar spin chirality on a triangular plaquette of sites $\bm{r}_i$, $\bm{r}_j$, and $\bm{r}_k$ with moments $\bm{S}_i$, $\bm{S}_j$, and $\bm{S}_k$ is defined as $\chi_s^{} = \bm{S}_i\,\cdot\,(\bm{S}_j\times\bm{S}_k)$. The effective magnetic field felt by a conduction electron passing over this plaquette is $\bm{b} \propto t\chi_s^{}\hat{\bm{n}}$, where $\hat{\bm{n}}$ is the plaquette normal vector and $t = t_{ij}t_{jk}t_{ki}$ is the hopping integral around the loop $i\rightarrow j\rightarrow k\rightarrow i$. The lattice of magnetic Co sites in \cns consists of two triangular sublattices. Thus, there are intra-sublattice plaquettes formed by three sites in the same basal plane, as well as inter-sublattice plaquettes formed by two sites from a given sublattice and one site from the opposite one. We denote the intra- and inter-sublattices contributions as $\chi_s^{\parallel}$ and $\chi_s^{\perp}$ respectively. For intra-sublattice plaquettes, 
\begin{equation}
    \chi_s^{\parallel}(\bm{r}) =
    \chi[\sin(\alpha+\beta) - \sin\alpha - \sin\beta]\sin\phi_n\cos^2{\phi_n}
\end{equation}
where $\bm{q} = (\alpha\beta 0)/2\pi$ is the modulation wavevector, $\Phi_{\bm{r}} = \bm{Q}_0\cdot\bm{r} + \phi_n$, and $\chi = \pm 1$ is the helix chirality. This vanishes everywhere if $\alpha=0$, $\beta=0$, or $\phi_n=0,\frac{\pi}{2},\pi,\frac{3\pi}{2}$. It is also independent of $\psi$, $\mu$, and $\nu$. We can see that $\chi_s^{\parallel}$ has a fixed magnitude at all positions but oscillates in sign along the direction of $\bm{Q}_0$. The sum of $\chi_s^{\parallel}$ for the two sublattices is shown in Fig.~\ref{fig:chi_IP}. Since the sites on each sublattice are offset along the direction of $\bm{Q}_0$, there is a phase difference in the contribution from each sublattice. This causes some interference, seen as the black regions of vanishing $\chi_s$.

$\chi_s^{\perp}$ does not vanish for any particular direction of $\bm{q}$, but vanishes as $q\rightarrow 0$. Similarly to $\chi_s^{\parallel}$, $\chi_s^{\perp}$ oscillates in sign along $\bm{Q}_0$, but also oscillates along $\bm{q}$. The amplitude of these oscillations depends on $\nu$, with no oscillations for $\nu=0$. The precise pattern of $\chi_s^{\perp}$ depends on the choice of phase differences $\Delta\phi$ and $\Delta\psi$. 


The pattern along $\bm{Q}_0$ depends on the sum $\Delta\phi\!+\!\Delta\psi$ and the pattern along $\bm{q}$ depends only on $\Delta\psi$. For $\Delta\phi\!+\!\Delta\psi\! =\! 0\degree$ or $360\degree$, $\chi_s^{\perp}$ forms a stripe pattern. For $\Delta\phi\!+\!\Delta\psi \!=\! 180\degree$ $\chi_s^{\perp}$ forms a checkerboard pattern. For $\Delta\psi \! = \! 0\degree$, $\chi_s^{\perp}$ oscillates about zero along $\bm{q}$. For $\Delta\psi \! = \! 180\degree$, $\chi_s$ oscillates in magnitude along $\bm{q}$ without passing through zero with an amplitude that increases with increasing $\nu$.

Although the magnitude of $\chi_s^{\perp}$ is much greater than that of $\chi_s^{\parallel}$, the relative magnitude of the effective field $\bm{b}$ is more relevant. We have already projected $\chi_s^{\perp}$ onto the $z$-component of the plaquette normal vectors, so we only need to consider the relative hopping integrals around each type of plaquette, $t_{\parallel} = t^3$ and $t_{\perp} = t'^2t$, where $t$ and $t'$ are the in- and out-of-plane hopping integrals along a single bond. Since the ratio of bond lengths $a'/a \approx 1.2$~\AA, we expect $t'<t$. Thus, the relative contribution of $\chi_s^{\perp}$ will be reduced by a factor of $t'^2/t^2$. However, the qualitative behavior of $\bm{b}$ remains unaffected. 

$\chi_s^{\parallel}$ forms stripes with propagation vector $\bm{Q}_0$ and no variation along $\bm{q}$. The stripes on each sublattice are $\pm 90\degree$ out of phase for $\Delta\phi=0\degree$ or $180\degree$. $\chi_s^{\perp}$ depends on $\nu$ and forms complex structures that depend on the choice of the phase variables. We apply the constraint $\Delta\psi\!=\!0\degree$ or $180\degree$ to consider four different combinations of phases that result in stripe or checkerboard like patterns of chirality. In all cases, the magnitudes of $\chi_s^{\parallel}$ and $\chi_s^{\perp}$ vanish as $q\rightarrow 0$. Thus, the incommensurate modulation is essential for providing a finite local scalar spin chirality in \cns{}.

Due to the opposite canting between sublattices, $\chi_s^{\perp}$ is much larger than $\chi_s^{\parallel}$ for the values of $\nu$ we have found. Although the relative magnitude of the intra- and inter-sublattice hopping integrals are unknown, the effective field produced by $\chi_s^{\perp}$ will dominate for all feasible values and the qualitative behavior is unchanged by $\chi_s^{\parallel}$. In order to visualize $\chi_s^{}$, we project $\chi_s^{\perp}$ onto the $z$ component of the plaquette normal vectors, shown in Fig. 4(d) and (e) of the main text for two different possibilities of the relative phases. In all four cases, the scalar chirality develops an intricate pattern, modulated along both $\bm{Q}_0$ and $\bm{q}$.

Such a staggered chirality cannot directly account for the AHE. However, it may give rise to nonlinear or nonreciprocal responses \cite{hayami2022nonreciprocal}, as recently observed \cite{mi2023order}. Furthermore, the magnetic structure we observed may generate a net chirality if the moment size is nonuniform, which can be described by relaxing the orthonormality constraint on the vectors $\hat{\bm{u}}$, $\hat{\bm{v}}$, and $\hat{\bm{w}}$.


\bibliography{bib}